\documentclass[
 reprint,
 amsmath,amssymb,
 aps,
]{revtex4-2}

\usepackage{graphicx}
\usepackage{dcolumn}
\usepackage{bm}
\usepackage{hyperref}

\usepackage{xcolor}
\usepackage{soul}
\usepackage{easyReview}

\begin{document}

\preprint{APS/123-QED}

\title{Scanning quantum correlation microscopy with few emitters}

\author{Jaret J. Vasquez-Lozano}
\email{jaret.vaslo@gmail.com}

\author{Shuo Li}
 \email{shuo.li3@rmit.edu.au}
\affiliation{
 ARC Centre of Excellence for Nanoscale BioPhotonics, School of Science, RMIT University, Melbourne 3001, Australia 
}

\author{Andrew D. Greentree}
 \email{andrew.greetree@rmit.edu.au}
\affiliation{
 ARC Centre of Excellence for Nanoscale BioPhotonics, School of Science, RMIT University, Melbourne 3001, Australia 
}
\date{\today}

\begin{abstract}
Optical superresolution microscopy is an important field, where nonlinear optical processes or prior information is used to defeat the classical diffraction limit of light. Quantum correlation microscopy uses photon arrival statistics from single photon emitters to aid in the determination of properties including the number of emitters and their relative brightness.  Here we model quantum correlation microscopy in the few emitter regime, i.e. around four single photon emitters below the diffraction limit.  We use the Akaike Information Criterion to determine the number of emitters and we vary the relative contributions of intensity to quantum correlation information to determine contribution that provides optimal imaging. Our results show diffraction unlimited performance and a change in localisation scaling behaviour dependant on emitter closeness.
\end{abstract}

\maketitle

\section{\label{sec:level1}Introduction}

The quest to gain a greater understanding of biological systems and systems at the atomic scale has motivated the discovery of new microscopy techniques to overcome the diffraction limit, such as STED and STORM \cite{Hemmer16,Klar,Rust,Hemmer12,Sauer,Tam,Diaspro}. 

Although optical superresolution techniques like STED are in principle diffraction unlimited, in practice, the achievable resolution in biological systems is limited by damage induced by the optical beams: phototoxicity \cite{Icha}.  This limits the utility of superresolution, especially in live cell imaging, with limitations on either resolution, imaging time, or both \cite{Tam,Diaspro}.  As discussed by Hemmer and Zapata \cite{Hemmer12}, the optical power limitations apply across all superresolution techniques, although tradeoffs to achieve particular goals are usually possible.  This motivates the exploration of techniques that use information that is often ignored, and in that context we have chosen to explore the quantum statistics of light emitted in microscopy.

The use of quantum correlations in superresolution microscopy was proposed in one of the first superresolution schemes \cite{Hell}. Such quantum techniques typically employ the Hanbury Brown and Twiss experiment (HBT) \cite{HBT} in place of more conventional classical photon collection. HBT uses single-photon detectors and cross-correlates the received signals to provide information about the number of single-photon emitters in the field of view. The information obtained via HBT measurements can then be combined with classical microscopy results to improve their resolution \cite{Classen,Bartel}. 

A general treatment of the role that antibunching plays in microscopy can be found in \cite{Schwartz12}, which also clearly shows the achievable improvement in resolution.  These results were demonstrated experimentally in \cite{Schwartz13}.  Inspired by this work, Gatto Monticone \textit{et al.} \cite{Gatto} demonstrated how superresolution could be achieved when using confocal measurements and HBT on a cluster of classically unresolvable fluorescence color centers in diamond.
These treatments used HBT (and sometimes higher order)  measurements to effectively reduce the width of the point spread function (PSF), where the confocal PSF is raised to the power of $k$ where $k$ is the order of the correlation used in imaging.  This gives rise to an effective standard deviation for the higher order PSF of $\sigma/\sqrt{k}$ where $\sigma$ is the standard deviation for the standard (classical) PSF. Several other recent advances in utilising quantum correlation microscopy include works that use various different methodologies to achieve superresolution for thermal optical sources \cite{Oppel,Tsang,Classen16}. Other recent works include superresolution imaging using third and fourth order correlations \cite{Pearce}, and quantum imaging of remote bodies \cite{Howard}. A review on several quantum imaging techniques can be found in Ref. \cite{Altmann}.

One surprising result from quantum correlation microscopy is that the information obtained from the HBT experiment is \emph{qualitatively} different from that obtained using conventional intensity measurements.  Worboys \textit{et al.} demonstrated that the HBT signal for two emitters reveals the relative brightness of those emitters \cite{Worboys}. This insight enabled the development of a new protocol, quantum trilateration, where two single photon-emitters with unknown relative brightness can be localised to arbitrary precision on the basis of data from three measurement locations.  Such a protocol would be impossible using conventional intensity measurements alone as the number of free parameters (five) exceeds the number of measurements (three).  By contrast, HBT and intensity measurements combined provides intrinsic brightness and relative brightness. This leads to six different measurement results, enabling the localisation of the two emitters. These results generalise to three dimensions \cite{LLM+2023}.

Here, we show that the scaling of our approach with respect to time converges to $1/\sqrt{t}$ after a certain turning point. Our observations show that this turning point is dependant on the minimum spacing of the emitters. This behaviour is shown for several cases of three to four emitter with increasing complexity such as increased background levels and unequal emitter brightness. We present a simple heuristic for obtaining the turning point location by interpolating to find the intersect of the scaling plots before and after the turning point.
Our results show that the optimal ratio of intensity and correlation information is of relatively little significance in cases with no background, provided that both information sources are used. In cases with background, there is a slight preference for higher contributions of correlation information.
Finally, we demonstrate how the Akaike Information Criteria (AIC) \cite{Aka1974} can be used to constrain the number of emitters, with the accuracy of the calculation of the ground truth number of emitters improving as measurement time increases. Additionally, we observe that the measurement time required for more accurate emitter number determination decreases at higher background levels.

This manuscript is organised as follows. We first describe the HBT setup, and demonstrate how to generate the expected outputs  using the second order correlation function. We show the time scaling laws of our approach for several random configurations of three or four emitters, as well as how different weightings of intensity and correlation information affects these results. We then show the evolution of AIC results with increasing measurement time. We repeat this analysis for emitter configurations of: equal brightness, equal brightness with uniform background, and unequal brightness with and without background.

\section{Theory}

\subsection{Simulating second-order quantum correlations}

\begin{figure}[tb!]
    \centering
    \includegraphics[width = 0.9\columnwidth]{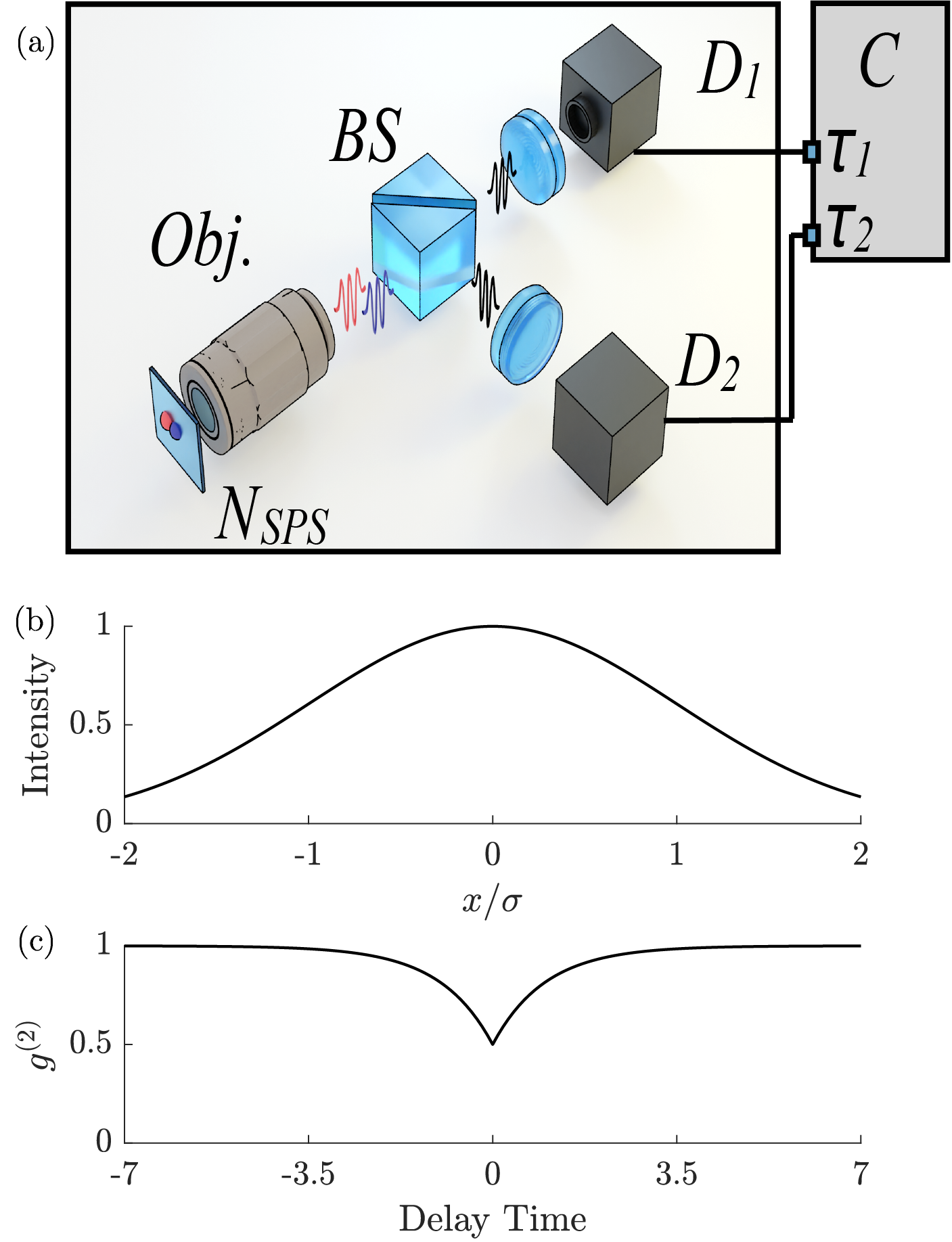}
    \caption{(a) Schematic of HBT experiment setup. Here, we are considering a confocal scan of some sample area ($N_{SPS}$) containing single photon emitters. The beam splitter and two detectors ($D_1$, $D_2$) monitored in coincidence allows for the study of photons from multiple emitters. Photon counts are correlated at $C$ to obtain the second order correlation function as a function of delay time, $\tau$. (b) Example cross section of normalised intensity profile obtained by summing photon counts collected at each measurement location. As the two emitters are close with respect to the diffraction limit, we expect to see a single peak. (c) Example second order correlation function ($g^{(2)}$) output obtained by correlating photon counts. The probability of all possible two photon detection events at detectors $D_1$ and $D_2$ for a given delay time are normalised by uncorrelated detection events $(\tau = \pm \infty)$. The example here is for the case of two equal brightness emitters which result in $g^{(2)}(\tau = 0) = 0.5$.}
    \label{fig:HBT}
\end{figure}

The second-order correlation function, or the Hanbury Brown and Twiss experiment, provides information about the probability of multi-photon emission from a particular field of view.  Figure~\ref{fig:HBT}(a) shows a schematic of the HBT setup that we are analysing. This two-detector setup measures the coincidence rate of photons for a system comprised of several single photon emitters.  The position of each emitter, $E_i$, is $\overrightarrow{x}_i = (x_i,y_i)$. Photons are collected from the sample via the microscope objective, which is modelled via the microscope point spread function (PSF), and the signal is split to two detectors via a beamsplitter. 
Detector signals can be analysed by summing, which retrieves the conventional intensity measurement, or in coincidence, which provides the HBT signal (second order correlation). Note that because the signals are electronic, both sets of data can be collected simultaneously.  
 
To perform imaging of a fluorescent sample with unknown number of emitters, we envisage scanning the detector across the field of view, as in \cite{Gatto}, although wide-field approaches are possible \cite{Schwartz12}, especially via the developments in single photon avalanche diode arrays \cite{Claudio, Zappa}. 

For $N$ single photon emitters, the second order correlation function measured by the HBT setup is \cite{Worboys}

\begin{equation}
    g^{(2)}_N = \frac{2\sum_{i=1}^{N-1}\sum_{j=i+1}^{N}P_i P_j}{\sum_{i=1}^{N}\sum_{j=1}^{N}P_i P_j}
\label{eq:1}
\end{equation}

\noindent where $P_i$ and $P_j$ are the detection probabilities of photons from the respective emitter, $i$ and $j$.  The detection probability is found from the product of the excitation probability, emission probability, PSF,  all optical losses in the microscope and any differences between the emitters (for example different orientations of the emitters \cite{Davin}). Note that although the second order correlation function is often written as a function of the time delay between detections, $\tau$,  for our purposes we are only  concerned with the correlations at zero time delay, ie when both detectors fire simultaneously. This corresponds to the the HBT signal at delay time zero, as shown in Figure~\ref{fig:HBT}.

As can be seen from Eqn.~\ref{eq:1}, the second order correlation provides information about the number of emitters, and provides maximum number discrimination when there are two emitters in the field of view. However, as shown by Li \textit{et al.} \cite{Shuo}, higher order correlations provide more information about the number of emitters as the number of emitters increases.  Nevertheless we stay with the second order correlation function as it is the most convenient and accessible experimental setup.

The general form for the second order correlation function in Eqn.~\ref{eq:1} provides one mechanism for treating background either arising from background fluorescence, detector dark counts, or both.  Here, we model the background as arising due to a large number of single photon emitters, where the probability that each will emit in any given time is low, however the product of this probability with the number of emitters in a diffraction limited spot is not negligible.  We further assume that the probability of emission/detection from each of these emitters is equal and that the density of emitters across the entire sample is constant.  We set the detection probability from each background emitter as $P_{bg} \ll P_{i}$, where $i$ here refers only to the emitters we wish to characterise (i.e. the bright, non-background emitters).  To determine the total number of emitters, we should integrate the background density over the point spread function, however in the limit that the number of emitters is large, we can approximate the background as arising from a large number emitters $\mathcal{N} \rightarrow \infty$ where $\mathcal{N}P_{bg} \lesssim \min({P_i})$  In this limit, the second order correlation becomes \cite{Worboys,Davin}

\begin{equation}
    g^{(2)}_{N} = \frac{2\left(\sum_{i=1}^{N-1}\sum_{j=i+1}^{N}P_i P_j + \sum_{i=1}^{N}\mathcal{N}P_{bg}P_i + \frac{\mathcal{N}^{2} P_{bg}^2}{2}\right)}{\sum_{i=1}^{N}\sum_{j=1}^{N}P_i P_j + 2\sum_{i=1}^{N}\mathcal{N}P_{bg}P_i + \mathcal{N}^2 P_{bg}^2}
\label{eq:g2BG}
\end{equation}

For simplicity we assume a Gaussian PSF, which is a good approximation to the more correct Airy function \cite{Stallinga}.  Hence the detection probability for the emitters is modelled via
\begin{equation}
    P_i(x,y) = P_{i,0} \exp[-\frac{(x-x_i)^{2} + (y - y_i)^2}{2 \sigma^2}],
\label{eq:G}
\end{equation}
where $(x_i,y_i)$ is the location of emitter $i$ and $(x,y)$ is the center of the microscope PSF. We have also introduced $P_{i,0}$, which we term the \emph{intrinsic} brightness of emitter $i$, which is defined as the probability of detecting a photon from the emitter when it is in the maximum of the PSF.

The standard deviation for the Gaussian PSF is related to the numerical aperture of the microscope via
\begin{align}
    \sigma \approx \frac{0.21 \lambda}{\text{NA}},
\end{align} 
for wavelength $\lambda$ and numerical aperture NA.  To avoid details of wavelength and numerical aperture, we normalise all of our distances and resolutions with $\sigma$.  

\begin{figure}[tb!]
    \centering
    \includegraphics[width = 0.9\columnwidth]{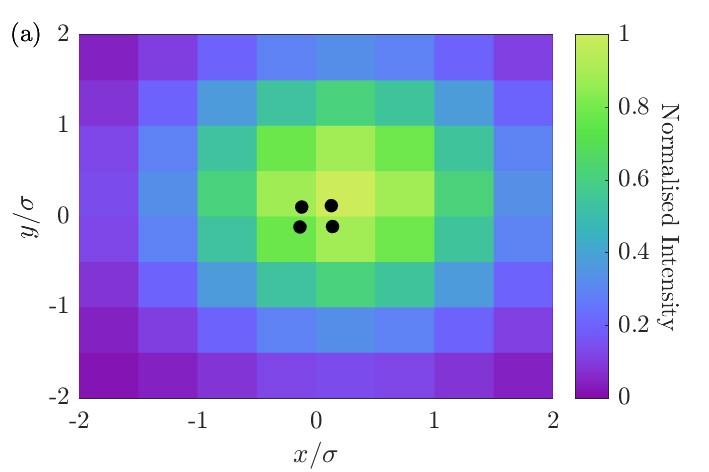}
    \includegraphics[width = 0.9\columnwidth]{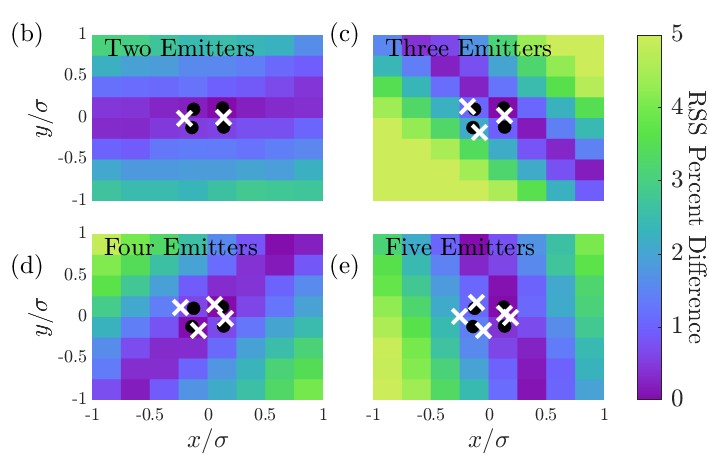}
    \caption{(a) Intensity image of $N = 4$ equal brightness emitters at locations shown by the black dots. The color axis is normalised to the maximum obtained brightness. (b)-(e) Four different fits to the data shown in (a) for 2 to 5 emitters (fitting results shown as crosses).  The color shows the percentage difference between the fitted brightness and the original signal brightness at each measurement location. While the difference in RSS values shows changes in the algorithm's preference for fitting, without prior knowledge of emitter number and locations any of these fits could be a potential candidate.}
    \label{fig:2}
\end{figure}

Figure~\ref{fig:2} (a) shows a simulation of a particular four emitter configuration in the long time limit, where each emitter has the same intrinsic brightness.  In the long time limit, the total number of received counts at location $(x,y)$ is proportional to $\sum_i P_i(x,y)$.  We imagine the detector system being scanned across our sample, in this case in a $9\times 9$ grid.  The  color axis shows the intensity (total detected photon counts) normalised by the maximum received number of counts. In Figure~\ref{fig:2}(b)-(e), we compare the ground truth case in Figure~\ref{fig:2} (a) to several different potential fits with different numbers of emitters. Based on the the percentage difference of the fitting Residual Sum of Squares (RSS), which vary within 1 percent in the area containing the emitters, and 5 percent near the bounds of the field of view, we show that on the basis of intensity-only that there are multiple acceptable fits.  This is a consequence of the well-known diffraction limit.

We are mostly concerned with the quantum imaging as a function of time, so as to best determine how the resolution varies in practical scenarios.  We therefore simulate the counts obtained using intensity and coincidence (HBT) via the MATLAB function \textsc{poissrnd} \cite{poissrnd}.

\begin{align}
    c_i &= \textsc{poissrnd}(P_{i}t), \nonumber \\
    c_{bg} &= \textsc{poissrnd}(\mathcal{N}P_{bg}t), \nonumber \\
    c_{i,j} &= \textsc{poissrnd}(P_{i}P_{j}t), \nonumber \\
     c_{i,bg} &= \textsc{poissrnd}(P_i\mathcal{N}P_{bg}t), \nonumber \\
    c_{bg,bg} &= \textsc{poissrnd}(\mathcal{N}^2P_{bg}^2t),
    \label{eq:4}
\end{align}
where $t$ is the total detection time (not the HBT correlation time shown Fig.~\ref{fig:HBT}), $c_i$ is the number of counts from emitter $i$, $c_{bg}$ is the number of counts from the background emitters, $c_{i,j}$ is the number of coincidences from emitters $i$ and $j$ for $i\neq j$, $c_{i,bg}$ is the number of coincidences between emitter $i$ and the background and we assume $c_{i,i}=0$. Because of the large number of background emitters, we cannot assume that the probability of coincidences solely from the background, $c_{bg,bg}$, is negligible.

For a system of $N$ emitters, with background and time dependence \cite{Davin}, Eq. \ref{eq:g2BG} becomes

\begin{align}
    g^{(2)}_{N}(t) = \frac{2\left(\sum_{i=1}^{N-1}\sum_{j=i+1}^{N} c_{i,j}  + \sum_{i=1}^{N}c_{i,bg} + \frac{c_{bg,bg}}{2}\right)}{\sum_{i=1}^N\sum_{j=1}^{N}c_{i,j} + 2\sum_{i=1}^{N}c_{i,bg} + c_{bg,bg}}
\label{eq:g2cBG}
\end{align}

The inclusion of background coincidence with $i,j$ emitters will lead to an increase the value of $g^{(2)}(0)$ in locations where typically, few coincidence events would be detected due to distance between emitters or low measurement times. Background counts correlating with other background counts will also need to be considered and will lead to $g^{(2)}(0)$ approaching 1 in areas where the $c_{i}$ are small.

\subsection{Calculation of effective PSF and relative Weighting of intensity and correlation information}

Using the time dependant second order correlation function, we are now able to generate the expected $g^{(2)}(0)$ for a given collection time. To obtain robust statistics for the expected uncertainty in localisation, we perform for 200 independent Monte-Carlo trials for each collection time.

Estimated emitter locations are obtained via minimising the least-square estimation of our intensity and correlation data. From Eqns.~\ref{eq:4}, individual photon counts, including $c_i$ and $c_{bg}$, provide the intensity data, $I$, while our coincident counts: $c_{i,j}$, $c_{i,bg}$, and $c_{bg,bg}$ provide the correlation data, $C$.

\begin{align}
    \text{RSS} = &\sum_{i=1}^{n}(I_{\text{data},i}-I_{\text{estimate},i})^{2} ... \nonumber \\
    &+\sum_{i=1}^{n}(C_{\text{data},i}-C_{\text{estimate},i})^{2}
\label{eq:10}.
\end{align}

where RSS refers to the residual sum of squares measure of the difference between the estimates and the Monte Carlo data.

Following Worboys \textit{et al.} \cite{Worboys}, we determine an estimate for the uncertainty by constructing an effective point spread function.  This is achieved by finding the centroid of the data and constructing a polygon that contains the 39.5\% of the points, i.e., that includes one standard deviation of the points.  The area of this polygon is then used to determine the diameter of an equivalent, circular region.  This diameter is then the effective width, $w_{\text{eff}}$ of the localisation precision, which is analogous to the more usual resolution width for microscopy. This approach also limits the influence that outliers caused by local minima in the RSS minimisation process has on the polygon size.

\begin{equation}
    w_\text{eff} = 2\sqrt{\frac{A_{39.5\%}}{\pi}}
\label{eq:9}.
\end{equation}

To determine the superresolution factor (improvement over the diffraction limit), $\gamma$, we take the average of the $w_\text{{eff}}$ for each emitter and divide by the confocal width, $2\sigma$, which is obtained from the experimental configuration.
 
An alternative to estimating the PSF achieved with quantum correlations in imaging applications is given by Ref.~\cite{Gatto}. There, expressions for higher order correlation functions are provided including coefficients that vary with order. $g^{(2)}(0)$ has a coefficient of 1, meaning that the approach used in Ref.~\cite{Gatto} uses intensity and $g^{(2)}(0)$ information equally. To investigate if the relative weighting between intensity and $g^{(2)}(0)$ affects $w_\text{{eff}}$, we introduced a parameter, $\alpha$, which is added to Eqn.~\ref{eq:10} to adjust the relative weighting.

\begin{align}
    \text{RSS} = (\alpha-1)&\sum_{i=1}^{n}(I_{\text{data},i}-I_{\text{estimate},i})^{2} ... \nonumber \\
    &+\alpha\sum_{i=1}^{n}(C_{\text{data},i}-C_{\text{estimate},i})^{2}
\label{eq:11}.
\end{align}
We explore $w_{\text{eff}}$ as a function of $\alpha$ to determine the extent to which $g^{(2)}(0)$ influences the localisation precision, where $\alpha = 0.5$ corresponds to the result in Ref.~\cite{Gatto}.  As will be seen in the Monte Carlo results below, providing $\alpha \neq 0~\text{or}~1$ (i.e. we are fitting using either intensity or $g^{(2)}(0)$ data only), $w_{\text{eff}}$ has only weak dependence on $\alpha$, as is expected for a minimisation process.

\subsection{Akaike Information Criteria for estimating emitter number}

The Akaike-Information-Criteria (AIC) is a mathematical tool used for determining goodness of fit \cite{Akaike}. The likelihood function of several models are compared to obtain a score, with a lower score signifying a greater likelihood of fit given the data. The scores are penalised based on the number of fitted parameters which mitigates overfitting.
As we are using least-square estimation in our fitting algorithm, we use a form of the AIC equation that employs the residuals of the fitting in the likelihood term \cite{Burnham}:

\begin{equation}
    \nu = 2k + n\ln(\hat{\sigma}^{2})
\label{eq:12}.
\end{equation}

\noindent where $\nu$ is the AIC score, $n$ is the sample size, $k$ is the number of fitted parameters, and $\hat{\sigma}$ is the reduced chi-square statistic which is defined as: $\text{RSS}/n$.

Here, the only relevant metric of the AIC score is the difference in AIC scores between models. Therefore, to compare models for their goodness of fit, we use the score the best model, $\nu_{\text{min}}$, and other candidate models, $\nu_i$, to calculate their relative goodness of fit\cite{Burnham}.

\begin{equation}
   L_{\text{AIC}} = \exp[(\nu_{\text{min}}-\nu_{i})/2],
\label{eq:13}
\end{equation}
where $L_{\text{AIC}}$ is the relative goodness of fit. This results in the best model having a $L_{\text{AIC}}$ of 1 (when $\nu_i = \nu_{\text{min}}$). Models with $L_{\text{AIC}}$ values close to zero have been determined by the algorithm to not accurately fit the data.

 \section{Study of Equal Brightness Emitter Configurations Without Background}

In practical situations, the number of emitters will usually not be known \emph{a priori}.  For this reason we have chosen to study several cases of localisation of few emitters, including three and four emitters.  In this section we concentrate on equal brightness emitters with no background.  The first case we study is a contrived case of four emitters in a line, which will allow us to examine the two scaling regimes, short time and long time.  We next consider three- and four-emitter configurations with the emitters spaced closely with respect to the diffraction limit.  

Equal emitter brightness in this context means equal \emph{intrinsic} brightness. This does not mean that each emitter will give rise to the same number of \emph{detected} photons, because each emitter is located in a different position relative to the collection point spread functions.  Instead this means that the maximum number of collected photons from each emitter would be the same if the emitters were located at the same positions relative to the collection point spread functions.

\subsection{Four emitters in a line}\label{sect:FourLine}

One of the primary challenges of a localisation technique is how it handles points that are close with respect to the diffraction limit. We expect that as emitter spacing decreases, the ability to resolve emitters will worsen. To study the effects of emitter spacing in our approach, we start by studying a specific case of four emitters in a line with an order of magnitude difference in spacing between two emitters. We will be calculating the improvement in $w_{\text{eff}}$ over time.

We consider four emitters of equal brightness, aligned diagonally in positions: $(x_1,y_1) =(0.010,0.060)\sigma$, $(x_2,y_2) = (0.025,0.075)\sigma$, $(x_3,y_3) = (0.25,0.3)\sigma$, and $(x_4,y_4) = (0.400,0.450)\sigma$. This results in two sets of emitters with emitter spacings of: $0.212\sigma$ and $0.0212\sigma$. These two sets of emitters are separated by a spacing of $0.318\sigma$.

\begin{figure}[tb!]
    \centering
    \includegraphics[width = 0.9\columnwidth]{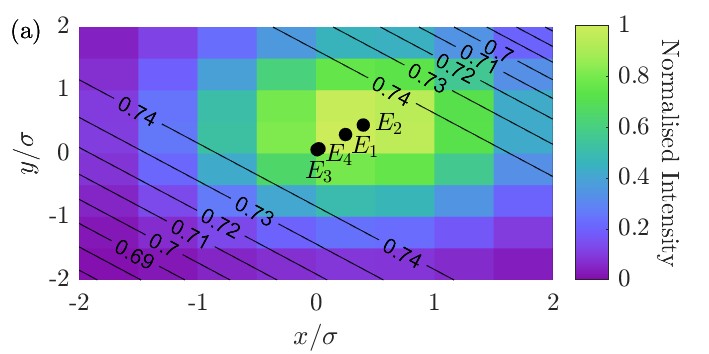}
    \includegraphics[width = 0.9\columnwidth]{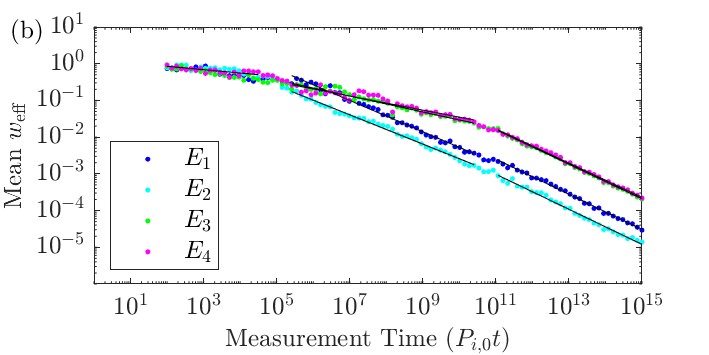}
    \caption{(a) Intensity plot and $g^{(2)}(0)$ contour of 4 emitters on a field with no background at positions: $(x_1,y_1) =(0.010,0.060)\sigma$, $(x_2,y_2) = (0.025,0.075)\sigma$, $(x_3,y_3) = (0.25,0.3)\sigma$, and $(x_4,y_4) = (0.400,0.450)\sigma$. Black dots indicate ground truth emitter position. (b) $w_{\text{eff}}$ scaling with measurement time. Fit-lines show where linear fits were obtained to determine scaling laws for each $w_{\text{eff}}$.} 
    \label{fig:Dline}
\end{figure}

The configuration is shown in Figure~\ref{fig:Dline} (a), with $w_{\text{eff}}$ as a function of measurement time shown in Figure~\ref{fig:Dline} (b). From this configuration, we can clearly see two distinct turning points for each $w_{\text{eff}}$ line, which we term `knees', where the gradient of $w_{\text{eff}}$ changes with time. This relationship between $w_{\text{eff}}$ and time can be expressed as \cite{Worboys}:

\begin{equation}
    w_{\text{eff},k} = 10^{c_{k}}t^{m_{k}}
\label{eq:weffM}.
\end{equation}
where $t$ is time, $c$ is the $y-$intercept on the logarithmic scale, and $m$ is the gradient of $w_{\text{eff}}$. $c$ and $m$ are determined by fitting $w_{\text{eff}}$ linearly between knees, with $k$ signifying prior to which knee the data is fitted (in order from left to right). As shown in \cite{Worboys}, $c$ is dependant on the number of measurement locations and the emitter configuration. As we are only interested in scaling laws, we will be focusing on the gradient, $m$.

The two close and two far emitters show knees at different measurement times, but with the same qualitatively behaviour. For the two emitters of spacing $0.212\sigma$, the scaling first changes from a gradient $m_1 = $ -0.09 $\pm$ 0.02, to $m_2 = $ -0.43 $\pm$ 0.02 and $m_2 = $ -0.40 $\pm$ 0.01 for emitters $P_{1,0}$, and $P_{2,0}$, respectively. It then transitions to a gradient of $m_3 = $ -0.48 $\pm$ 0.01 and $m_3 = $ -0.48 $\pm$ 0.02 for emitters $P_{1,0}$, and $P_{2,0}$, respectively. For the two emitters of spacing $0.0212\sigma$, the scaling first changes from a gradient of $m_1 = $ -0.09 $\pm$ 0.02, to $m_2 = $ -0.22 $\pm$ 0.02 and $m_2 = $ -0.19 $\pm$ 0.02 for emitters $P_{3,0}$, and $P_{4,0}$, respectively. It then transitions to a gradient of $m_3 = $ -0.471 $\pm$ 0.009 and $m_3 = $ -0.47 $\pm$ 0.01 for emitters $P_{3,0}$, and $P_{4,0}$, respectively. These results are summarised in Table~\ref{Dlinetable}.

\begin{table*}[tb!]
\caption{Summary of configuration shown in Fig \ref{fig:Dline}. Emitters are categorised by the distance to the closest other emitter in order to create two sets of emitters with a spacing of one order of magnitude difference. The gradients of the fits obtained from three different linear fits for each emitter's $w_{\text{eff}}$ is given. Gradients $m_1$, $m_2$, and $m_3$ correspond to fits obtained from approximately times: $10^{2}$ to $10^{5}$, $10^{5}$ to $10^{11}$, and $10^{11}$ to $10^{15}$, respectively.}
\begin{center}
\begin{tabular}{| c | p{2.8cm} | p{2.8cm} | p{2.8cm} | p{2.8cm} |}
\hline
\hline
 Emitter & $(x_i,y_i)/\sigma$ & $m_1$ & $m_2$ & $m_3$ \\
 \hline \hline
$E_1$ & $(0.01,0.06)$ & -0.09 $\pm$ 0.02  & -0.43 $\pm$ 0.02 & -0.48 $\pm$ 0.01\\ \hline
$E_2$ & $(0.025,0.075)$ & -0.09 $\pm$ 0.02 & -0.40 $\pm$ 0.01 & -0.48 $\pm$ 0.02 \\ \hline
$E_3$ & $(0.25,0.3)$ & -0.09 $\pm$ 0.02 & -0.22 $\pm$ 0.02 &  
-0.471 $\pm$ 0.009\\ \hline
$E_4$ & $(0.4,0.45)$ & -0.09 $\pm$ 0.02 & -0.19 $\pm$ 0.02 & 
-0.47 $\pm$ 0.01\\ \hline
\hline
\end{tabular}
\end{center}
\label{Dlinetable}
\end{table*}

By observing where we see knees in our $w_{\text{eff}}$ slopes, we find that that knees occur when $w_{\text{eff}}$ becomes less than the spacing between the two emitter sets, $0.318\sigma$, and the two closest emitters $0.0212\sigma$. We expect that if the spacing between the emitter sets and the two far emitters was greater we would see an additional knee there. From these results, we see a clear link between emitter spacing and scaling behaviour. We can characterise the scaling of $w_{\text{eff}}$ based on the time where our turning point occurs, which we can denote as $t_\text{knee}$. A simple means to obtain $t_\text{knee}$ would be to interpolate between the $w_{\text{eff}}$ before and after $t_\text{knee}$, i.e.,

\begin{equation}
w_{\text{eff}}(t) = \min(10^{c_1} t^{m_1},10^{c_2} t^{m_2})
\label{eq:weffKnee}.
\end{equation}
Experimentally, we would expect that the emitters would be more evenly distributed among the field of view. This would result in only one knee being easily discernible, which would result in only two $w_{\text{eff}}$: $w_{\text{eff},1}$ and $w_{\text{eff},2}$. For simplicity, we will only consider such cases in this paper moving forward. Thus, we will be obtaining two gradients, $m_1$ \& $m_2$,
and single $t_\text{knee}$ which we expect will be where $w_{\text{eff}}$ becomes less than the minimum spacing of the emitters, $d_\text{min}$. As stated we will obtain $t_\text{knee}$ by using Eqn.~\ref{eq:weffKnee}.

\subsection{Three emitters of equal brightness with no background}

As we are interested in the localisation of more than two emitters, the simplest case we consider is a configuration of three equal brightness, unresolved emitters.

\begin{figure}[tb!]
    \centering
    \includegraphics[width = 0.9\columnwidth]{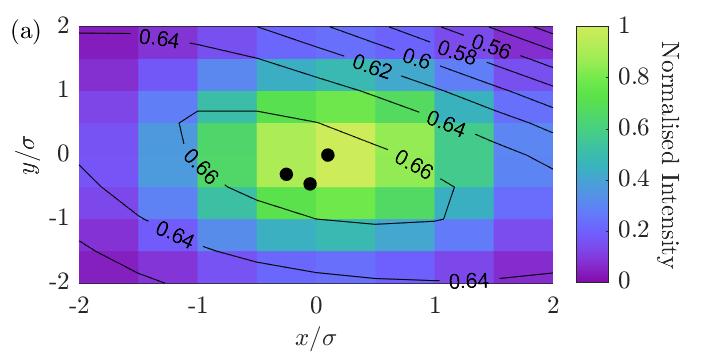}
    \includegraphics[width = 0.9\columnwidth]{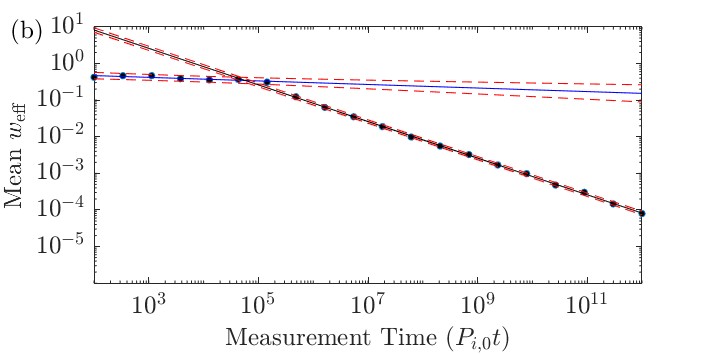}
    \includegraphics[width = 0.9\columnwidth]{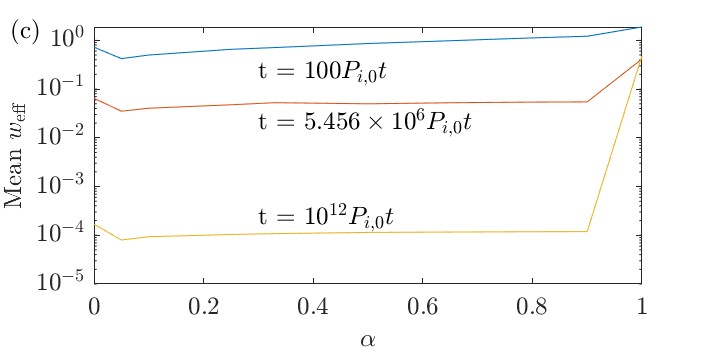}
    \includegraphics[width = 0.9\columnwidth]{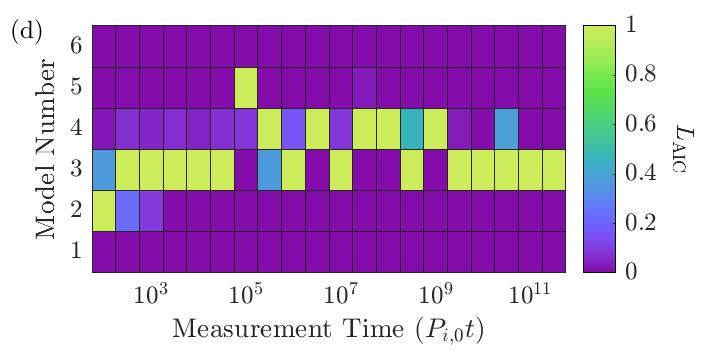}
    \caption{(a) Intensity plot and $g^{(2)}(0)$ contour of three emitters on a field with no background at positions: $(x_1,y1) =(-0.25,-0.3)\sigma$, $(x_2,y_2) = (-0.05,-0.45)\sigma$, and $(x_3,y_3) = (0.1,0)\sigma$. Black dots indicate ground truth emitter position. (b) $w_{\text{eff}}$ scaling with measurement time. Blue line shows fitted data belonging to $w_{\text{eff},1}$ used to obtain $m_1$. Black line shows fitted data belonging to $w_{\text{eff},2}$ used to obtain $m_2$. Red dashed lines show $95\%$ confidence interval of fits. The lines are extrapolated past the fitted data, showing the projected trajectory of $w_{\text{eff}}$ if the scaling behaviour did not change. $t_\text{knee}$ is interpolated to be $6.7 \times 10^{4} \pm 7 \times 10^{3}P_{i,0}t$. (c) Average of all emitter $w_{\text{eff}}$ achieved with different intensity to $g^{(2)}(0)$ weighting ratio, $\alpha$. An improvement in $w_{\text{eff}}$ with a weighting between 0 and 1 means that the algorithm performs best using both data sources (d) AIC goodness of fit given the data for models with differing numbers of emitters. Best model has a score of 1, with others having a score showing the goodness of fit relative to the optimal model.}
    \label{fig:cfig4}
\end{figure}

Figure~\ref{fig:cfig4} (a) shows an intensity map of three, equal brightness emitters on a field with no background that would be obtained after an infinitely long measurement time. The emitters are located at positions: $(x_1,y1) =(-0.25,-0.3)\sigma$, $(x_2,y_2) = (-0.05,-0.45)\sigma$, and $(x_3,y_3) = (0.1,0)\sigma$.

Figure~\ref{fig:cfig4} (b) shows $w_{\text{eff}}$ as a function of measurement time. We see two distinct gradients in the $w_{\text{eff}}$ line which we can identify as $m_1$ and $m_2$, as described in Section~\ref{sect:FourLine}. The $w_{\text{eff}}$ gradient begins with $m_1 = -0.05 \pm 0.03$ until $t_\text{knee} = 6.7 \times 10^{4} \pm 7 \times 10^{3}P_{i,0}t$ where it transitions to $m_2 = -0.500 \pm 0.007$. This transition occurs where $w_{\text{eff}}$ approximately becomes less than $d_{\text{min}}$, which is $0.250\sigma$ for this configuration, in keeping with the intuition built up in Section~\ref{sect:FourLine}.

To explore how intensity and $g^{(2)}(0)$ combine to provide higher resolution, in Figure~\ref{fig:cfig4} (c), we show $w_{\text{eff}}$ as the weighting parameter $\alpha$ from the residual sum of squares (eq.~\ref{eq:10}) is varied. In the limit $\alpha=0$ the fit uses only intensity information for emitter localisation, and $\alpha = 1$ corresponds to only $g^{(2)}(0)$ information for localisation.

We observe that any value of $\alpha$ that is not 0 or 1 results in better localisation than using only one information source. The variation in $w_{\text{eff}}$ is relatively minor resulting in a relatively flat plateau for $0 < \alpha < 1$, where random fluctuations in the algorithm's localisation and the exact emitter configuration plays a role in determining the flatness. The flatness of the $w_{\text{eff}}$ suggests that the exact contribution weighting matters little as long as both intensity and correlation information is used. Simply using a value of $\alpha = 0.5$ so that equal weightings are used would be appropriate. This is analogous to the approach used in Ref.~\cite{Gatto}, where equal contributions are used. In the following cases, we will provide the value of $\alpha$ that provides the minimized $w_{\text{eff}}$ value (i.e, the lowest $w_{\text{eff}}$ in the $\alpha$ plateau).

With optimal weighting to minimize $w_{\text{eff}}$, we obtain a superresolution factor, $\gamma = w_{\text{eff}}/(2\sigma)$, of 9 $\pm$ 3 at $\alpha = 0.05$, $\gamma$ = 114 $\pm$ 4 at $\alpha = 0.05$, and $\gamma$ = 5.0 $\times 10^{4}$ $\pm$ 600 at $\alpha = 0.05$, for times $100P_{i,0}t$, $5.456 \times 10^{6}P_{i,0}t$, and $10^{12}P_{i,0}t$, respectively.

Figure~\ref{fig:cfig4} (d) shows the AIC goodness of fit given the data as a function of time. We see from the results that the model resulting in $L_{\text{AIC}} = 1$ (which indicates the model with the best estimation) changes depending on time. At low time, when the data is particularly noisy, $L_{\text{AIC}} = 1$ for the model containing two emitters. For this case, at approximately $10^{3}P_{i,0}t$, the $g^{(2)}(0)$ counts have increased such that the data is less noisy and we consistently see the three emitter model as the optimal model, consistent with the ground truth. However, due to fluctuations caused by the Poisson statistics of the emissions, the optimal model changes between three and four emitter models. This fluctuation in optimal models stops as we reach high measurement times such as $10^{10}P_{i,0}t$.

\subsection{Four emitters of equal brightness with no background}

\begin{figure}[tb!]
    \centering
    \includegraphics[width = 0.9\columnwidth]{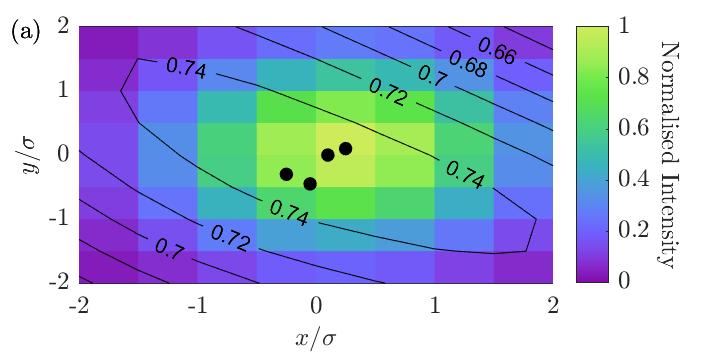}
    \includegraphics[width = 0.9\columnwidth]{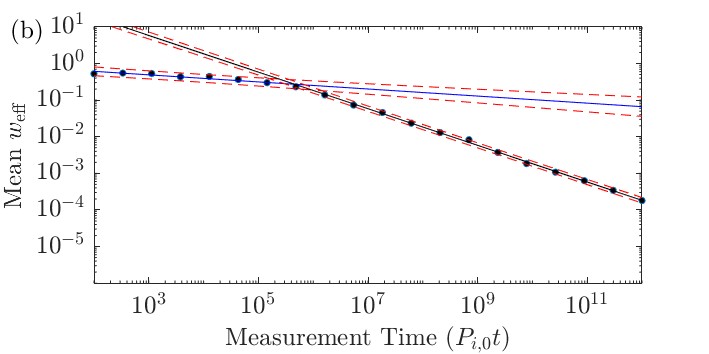}
    \includegraphics[width = 0.9\columnwidth]{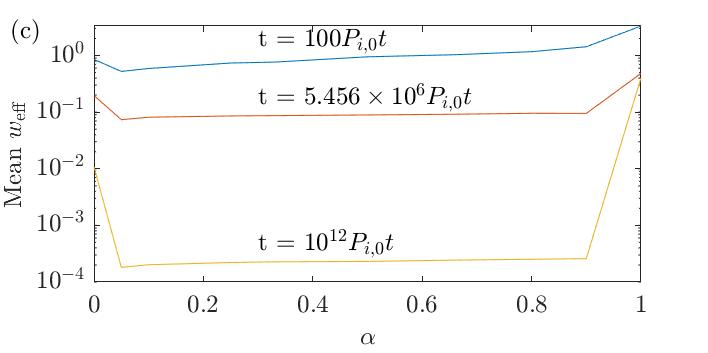}
    \includegraphics[width = 0.9\columnwidth]{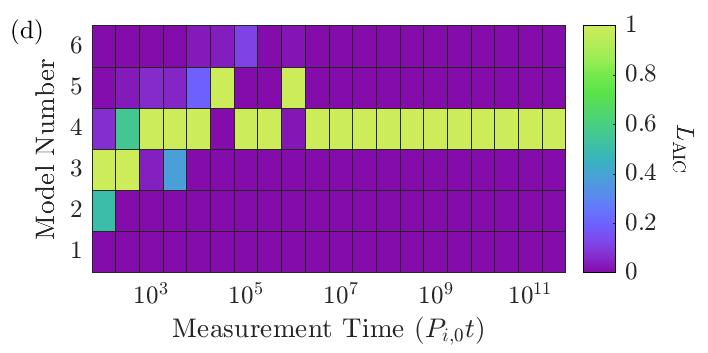}
    \caption{(a) Intensity plot and $g^{(2)}(0)$ contour of four emitters on a field with no background at positions: $(x_1,y1) =(-0.25,-0.3)\sigma$, $(x_2,y_2) = (-0.05,-0.45)\sigma$, $(x_3,y_3) = (0.1,0)\sigma$, and $(x_4,y_4) = (0.25,0.1)\sigma$. Black dots indicate ground truth emitter position. (b) $w_{\text{eff}}$ scaling with measurement time. Blue line shows fitted data belonging to $w_{\text{eff},1}$ used to obtain $m_1$. Black line shows fitted data belonging to $w_{\text{eff},2}$ used to obtain $m_2$. Red dashed lines show $95\%$ confidence interval of fits. $t_\text{knee}$ is $4.7 \times 10^{5} \pm 6 \times 10^{4} P_{i,0}t$. (c) Average of all emitter $w_{\text{eff}}$ achieved with different intensity to $g^{(2)}(0)$ weighting ratio, $\alpha$. As in Figure~\ref{fig:cfig4}, we see that $w_{\text{eff}}$ performs best when using values of $\alpha$ between 0 and 1. (d) AIC goodness of fit given the data for models with differing numbers of emitters. We see that at time approximately corresponding to $t_{\text{knee}}$, $L_{\text{AIC}} = 1$ consistently for the four emitter model.}
    \label{fig:cfig2}
\end{figure}

We now introduce an extra emitter into the problem to understand how the problem grows with increasing complexity. Figure~\ref{fig:cfig2} (a) shows the intensity map of 4 equal brightness emitters on a field with no background that would be obtained after an infinitely long measurement time. Overlayed onto the intensity map is a series of contours corresponding to $g^{(2)}(0)$. The emitters are located at positions: $(x_1,y_1) =(-0.25,-0.3)\sigma$, $(x_2,y_2) = (-0.05,-0.45)\sigma$, $(x_3,y_3) = (0.1,0)\sigma$, and $(x_4,y_4) = (0.25,0.1\sigma)$.

Figure~\ref{fig:cfig2} (b) shows the $w_{\text{eff}}$ scaling with respect to measurement time. The resolution scaling begins with $m_1 =  -0.10 \pm 0.04$ until approximately $t_{\text{knee}} = 4.7 \times 10^{5} \pm 6 \times 10^{4} P_{i,0}t$, where we transition to $m_2 = -0.50 \pm 0.01$.  As in the previous case, this transitions corresponds to when the mean $w_{\text{eff}}$ approximately becomes less than $d_{\text{min}}$ which is $0.1803\sigma$ for this configuration.

In Figure~\ref{fig:cfig2} (c), we show the calculated $w_{\text{eff}}$ with varied weighting parameter, $\alpha$. With optimal weighting to minimize $w_{\text{eff}}$, we obtain a superresolution factor of $\gamma$ = 8 $\pm$ 4 at $\alpha = 0.05$, $\gamma$ = 54 $\pm$ 4 at $\alpha = 0.05$, and $\gamma$ = 2.2 $\times 10^{4}$ $\pm$ 400 at $\alpha = 0.05$, for times $100P_{i,0}t$, $5.456 \times 10^{6}P_{i,0}t$, and $10^{12}P_{i,0}t$, respectively. 

Figure~\ref{fig:cfig2} (d) shows the AIC goodness of fit given the data as a function of time. As in the previous case, we see that the model resulting in $L_{\text{AIC}} = 1$ changes depending on time, achieving better fits with models at fewer emitters at low time. For this configuration, once we reach a measurement times of approximately $10^{3}P_{i,0}t$ we begin to consistently see the four emitter model as the optimal model, but we still see some variation in optimal fit until $5.456 \times 10^{6}P_{i,0}t$.

\subsection{Effective PSF from intensity only, and relationship between scaling and minimum emitter spacing}\label{sect:EQscale}

\begin{figure}[tb!]
    \centering
    \includegraphics[width = 0.9\columnwidth]{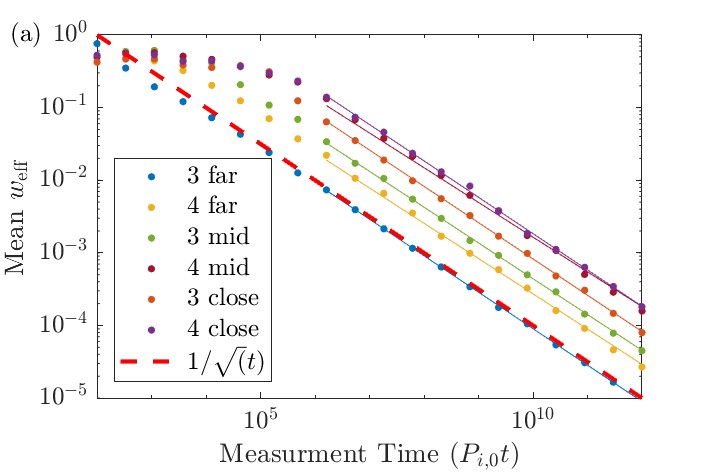}
    \includegraphics[width = 0.9\columnwidth]{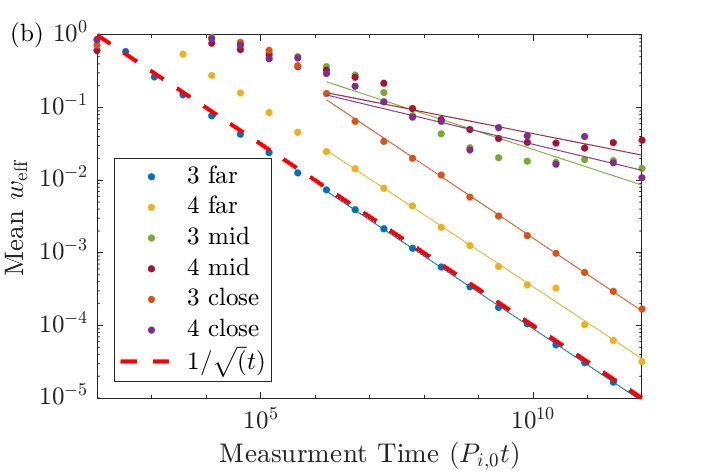}
    \caption{(a) Localisation scaling with time of six different emitter configurations categorised by their minimum emitter spacing relative to other configurations with the same emitter number, with `close' having the smallest spacing, and `far' being the furthest. (b) Resolution scaling for the same configurations in (a), using only intensity information. From top to bottom according to the legend, the minimum spacing for each configuration is: $1.75\sigma$, $0.81\sigma$, $0.52\sigma$, $0.40\sigma$, $0.25\sigma$, and $0.18\sigma$. Red-dashed line indicates $1/\sqrt{t}$ $w_{\text{eff}}$ scaling, which is the best expected scaling result.} 
    \label{fig:EQscale}
\end{figure}

Figure~\ref{fig:EQscale} shows a comparison of the resolution scaling with respect to measurement time for $g^{(2)}(0)$-and-intensity fitting (a) compared with intensity-only fitting (b), for six cases. The cases studied are summarised in Table~\ref{EQtable}. The cases have been identified based on their minimum emitter spacing relative to other configurations with the same number of emitters. We show $m_1$ and $m_2$ for all cases, and compare between the $g^{(2)}(0)$-with-intensity and intensity only cases, which may not necessarily be approximately $1/\sqrt{t}$ as we do not expect intensity only to be diffraction unlimited.

\begin{table*}[tb]
\caption{Summary of configurations shown in Figure~\ref{fig:EQscale}. Configurations are categorised as `far', `mid', or `close' based on their minimum emitter spacing relative to other configurations with the same number of emitters, $N$. Results obtained using $g^{(2)}(0)$-with-intensity $(I$ \& $C)$ and intensity-only $(I)$ fitting are compared. Results are in order of the emitter's minimum spacing, $d_\text{min}$. For cases where the interpolated $t_{\text{knee}}$ is $< 0$, the first data point where $w_{\text{weff}} < d_{\text{min}}$ is used and marked with *. If $w_{\text{weff}}$ is $< d_{\text{min}}$ from the first data point, or $m_1$ can not be obtained accurately due to too few data points, N/A is used for the corresponding data.}
\begin{center}
\begin{tabular}{|c|c| p{2cm} | p{2cm} | p{2cm} | p{2cm} | p{2cm} | c | p{2cm} | p{2cm} |}
\hline
\hline
$N$ & Separation & $(x_{i},y_{i})/\sigma$ & $m_1$ $(I$ \& $C)$ & $m_2$ $(I$ \& $C)$ & $m_1$ $(I)$ & $m_2$ $(I)$ & $d_{\text{min}}~[\sigma]$ & $t_\text{knee}~[P_{i,0}t$] $(I$ \& $C)$ & $t_\text{knee}~[P_{i,0}t$] ($I$) \\
 \hline \hline
3 & far& $(-1.3,1.0)$, $(-1.5,-0.8)$, $(0.2,-0.4)$ & N/A & -0.496 $\pm$ 0.005 & N/A & -0.500 $\pm$ 0.002 & 1.75 & N/A & N/A\\ \hline
4 & far & $(-1.1,0.4)$, $(-1.2,-0.6)$, $(0.2,-0.1)$, $(0.1,0.7)$ & N/A & -0.492 $\pm$ 0.007 & -0.1 $\pm$ 0.2 & -0.50 $\pm$ 0.01 & 0.81 & 1.128 $\times$ $10^{3}$ * & 400 $\pm$ 300\\ \hline
3 & mid & $(-0.4,0.2)$, $(-0.2,-0.45)$, $(0.1,0.05)$ & 0.1 $\pm$ 0.4 & -0.49 $\pm$ 0.01 & -0.12 $\pm$ 0.4 & -0.506 $\pm$ 0.007 & 0.52 & 4 $\times$ $10^{3}$ $\pm$ 3 $\times$ $10^{3}$ & 4.8 $\times$ $10^{4}$ $\pm$ 7 $\times$ $10^{3}$\\ \hline
4 & mid & $(-0.45,-0.3)$, $(-0.3,0.1)$, $(0.15,0.1)$, $(-0.05,-0.25)$ & -0.02 $\pm$ 0.08 & -0.48 $\pm$ 0.02 & -0.09 $\pm$ 0.08 & -0.25 $\pm$ 0.07 & 0.40 & 8 $\times$ $10^{4}$ $\pm$ 2 $\times$ $10^{4}$ & 8 $\times$ $10^{3}$ $\pm$ 1 $\times$ $10^{4}$ \\ \hline
3 & close & $(-0.25,-0.3)$, $(-0.05,-0.45)$, $(0.1,0)$ & -0.05 $\pm$ 0.03 & -0.500 $\pm$ 0.007 & -0.1 $\pm$ 0.1 & -0.50 $\pm$ 0.01 & 0.25 & 6.7 $\times$ $10^{4}$ $\pm$ 7 $\times$ $10^{3}$ & 7 $\times$ $10^{4}$ $\pm$ 2 $\times$ $10^{4}$\\ \hline
4 & close & $(-0.25,-0.3)$, $(-0.05,-0.45)$, $(0.1,0)$, $(0.25,0.1)$ & -0.10 $\pm$ 0.04 & -0.50 $\pm$ 0.01 & -0.15 $\pm$ 0.06 & -0.18 $\pm$ 0.08 & 0.18 & 4.7 $\times$ $10^{5}$ $\pm$ 6 $\times$ $10^{4}$ & 1.833 $\times$ $10^{7}$ *\\
\hline
\hline
\end{tabular}
\end{center}
\label{EQtable}
\end{table*}

We see in each of our cases that there is an improvement in emitter localisation when using $g^{(2)}(0)$ and intensity compared to using intensity alone. After a sufficiently long time ($t_{\text{knee}}$), we are able to achieve $1/\sqrt{t}$ $w_{\text{eff}}$ scaling for all cases when using $g^{(2)}(0)$. When using intensity-only, this is not achieved in the long time limit for the four emitter 'mid' and 'close' cases. While the $t_{\text{knee}}$ for the intensity-only $w_{\text{eff}}$ appears comparable to $g^{(2)}(0)$ and intensity at times, there is a larger relative uncertainty in $t_{\text{knee}}$ for intensity only cases due to uncertainty in fits. Generally, $t_{\text{knee}}$ for the intensity-only $w_{\text{eff}}$ is higher than for the $g^{(2)}(0)$-with-intensity $t_{\text{knee}}$.
 
We see in all cases, with the exception of the already resolved three emitter `far' case, that the data forms a knee where the $w_{\text{eff}}$ gradient converges to $1/\sqrt{t}$. As we have seen in the previous equal brightness cases, this knee corresponds to the point in the data where $w_{\text{eff}}$ becomes less than $d_{\text{min}}$. At this point, $w_{\text{eff}}$ is sufficiently small so that each emitter $w_{\text{eff}}$ is distinguishable, meaning the emitters are localised and we observe the optimal $w_{\text{eff}}$ scaling of $1/\sqrt{t}$. In the low time region prior to the knee, $w_{\text{eff}}$ improvement in time is slow as noisy intensity data and limited $g^{(2)}(0)$ counts lead to overlapping $w_{\text{eff}}$, creating ambiguity in both emitter locations and numbers. 

In summary, we observe that $t_{\text{knee}}$ is lower for cases with a larger $d_{\text{min}}$, which is in keeping with the heuristic that the $1/\sqrt{t}$ localisation scaling takes over when the emitters are independently resolved. The relationship is seen in Table~\ref{EQtable} and Figure~\ref{fig:EQscale}, where $t_{\text{knee}}$ increases as $d_{\text{min}}$ decreases.

When comparing cases in Figure~\ref{fig:EQscale} (a),
we see that $w_{\text{eff}}$ at a given time can be larger for four emitter cases than in three emitter cases of similar and sometimes smaller $d_{\text{min}}$. We expect that for some configurations, localisation of four emitters will not be as good as comparable three emitter cases, as there is loss in $g^{(2)}(0)$ sensitivity as the number of emitters increases, as studied in \cite{Shuo}.

\section{Study of equal brightness emitter configurations with background}

Background signals are present in all practical systems to some level.  These can be due to effects such as surface impurities or even dark count rates in detectors.  As mentioned above, we treat all sources of background signal equivalently.  For simplicity we will always assume that the background signal is constant across the field of view.  This approximation may not always be correct, for example in Ref.~\cite{HGG2017}, where the fabrication process led to an increase in background in the vicinity of the emitters.  Nevertheless, this approximation will assist in building insight into the role background plays in imaging configurations.

\begin{figure}[tb!]
    \centering
    \includegraphics[width = 0.9\columnwidth]{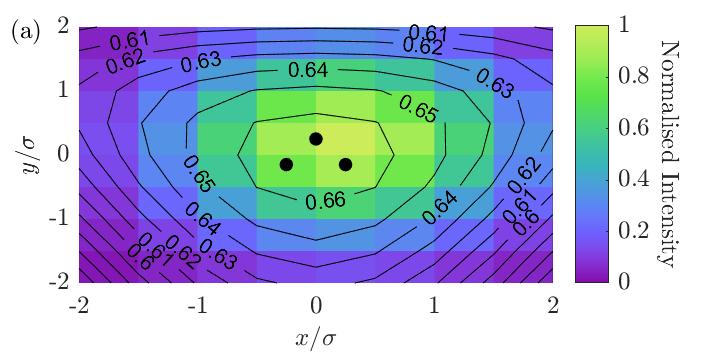}
    \includegraphics[width = 0.9\columnwidth]{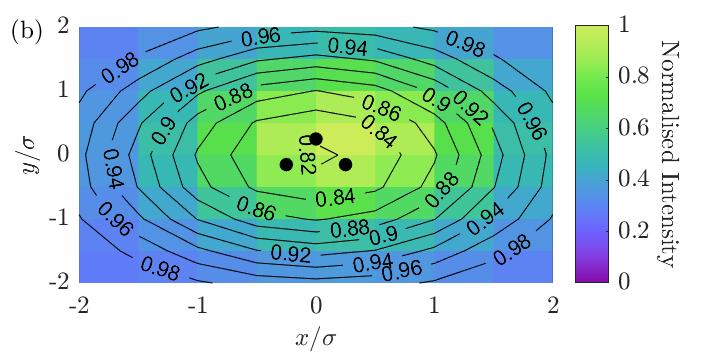}
    \caption{(a) Normalised intensity (color) and  $g^{(2)}(0)$ contours for a configuration of 3 equal brightness emitters on a field with (a) no background and (b)  uniform background of $0.2P_{i,0}$.  In each case the emitter locations (black dots) were $(x_1,y_1) = (0,0.25)$, $(x_2,y_2) = (-0.25,-0.125)$, and $(x_3,y_3) = (0.25,-0.125)$.  Background increases $g^{(2)}(0)$ at all locations, and  increases asymptotically to $g^{(2)}(0) = 1$ away from the emitters.}
    \label{fig:BGcomp}
\end{figure}

One important result from Eqns.~\ref{eq:g2BG} and \ref{eq:g2cBG} is that background-emitter correlations lead to HBT signals that would not otherwise be present in the no-background case. How this changes the characteristics of the $g^{(2)}(0)$ contour from what is observed without background can be seen in Fig.~\ref{fig:BGcomp}, where we compare the same configuration with a background of 0 and $P_{bg} = 0.2P_{i,0}$. These additional background-emitter correlations lead to an improvement in the $\alpha = 1$ results for $w_{\text{eff}}$, as will be shown below.

\subsection{Three emitters of equal brightness with constant background}

\begin{figure}[tb!]
    \centering
    \includegraphics[width = 0.9\columnwidth]{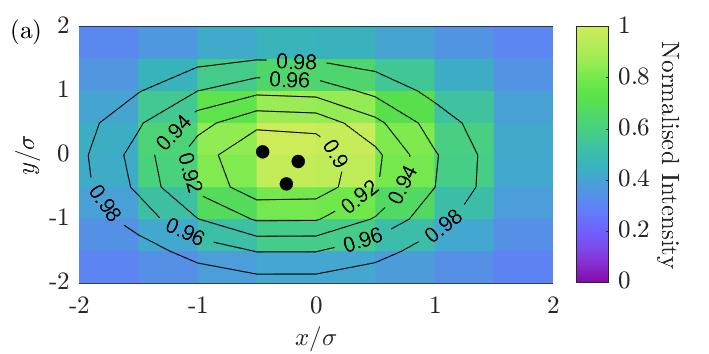}
    \includegraphics[width = 0.9\columnwidth]{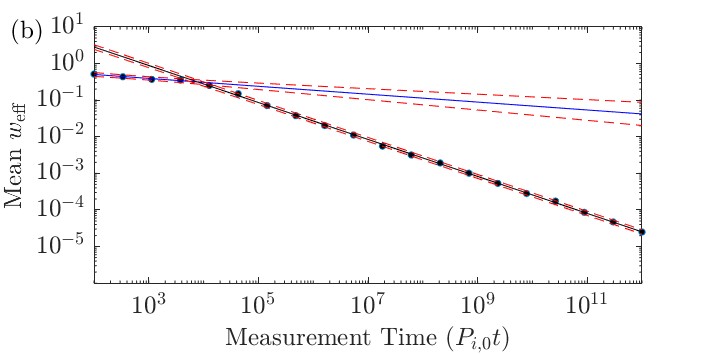}
    \includegraphics[width = 0.9\columnwidth]{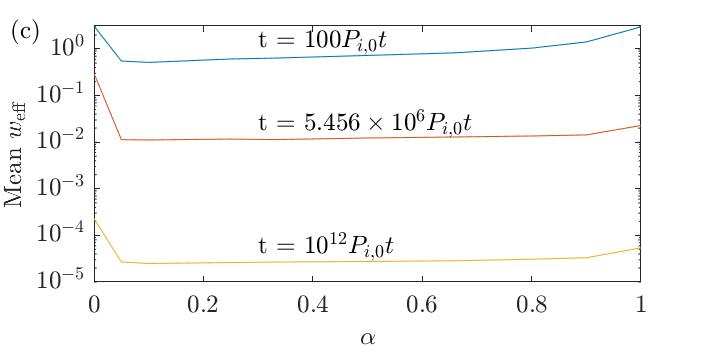}
    \includegraphics[width = 0.9\columnwidth]{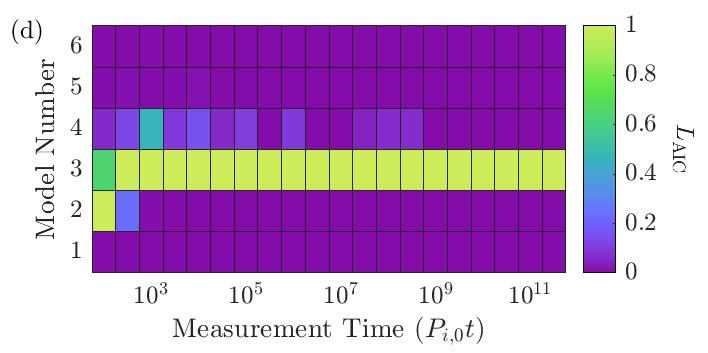}
    \caption{(a) Intensity (color) and $g^{(2)}(0)$ contour for 3 emitters with background $0.2P_{i,0}$ at positions: $(x_1,y_1) =(-0.45,0.05)\sigma$, $(x_2,y_2) = (-0.25,-0.45)\sigma$, and $(x_3,y_3) = (-0.15,-0.1)\sigma$. Black dots indicate ground truth emitter position. (b) $w_{\text{eff}}$ scaling with measurement time. Blue line shows fitted data belonging to $w_{\text{eff},1}$ used to obtain $m_1$. Black line shows fitted data belonging to $w_{\text{eff},2}$ used to obtain $m_2$. Red dashed lines show $95\%$ confidence interval of fits. $t_\text{knee}$ is $9 \times 10^{3} \pm 2 \times 10^{3} P_{i,0}t$. (c) Mean $w_{\text{eff}}$ achieved with different intensity to $g^{(2)}(0)$ weighting ratios, $\alpha$. Compared to the three emitter, no background case (Fig~.\ref{fig:cfig4} (c)), higher weightings of $g^{(2)}(0)$ and $g^{(2)}(0)$ alone have an improved localisation performance. (d) AIC likelihood of fit for models with differing numbers of emitters. Compared to the three emitter, no background case (Fig~.\ref{fig:cfig4} (d)), there is a reduction in time required for the $L_{\text{AIC}} = 1$ to match the ground truth. Additionally, there is less variation in optimal models as time increases.}
    \label{fig:cfig1bg}
\end{figure}

We begin our background analysis with three, equal brightness emitters in a field with a background of $0.2P_{i,0}$. The emitters are located at positions: $(x_1,y1) =(-0.45,0.05)\sigma$, $(x_2,y_2) = (-0.25,-0.45)\sigma$, and $(x_3,y_3) = (-0.15,-0.1)\sigma$. Figure~\ref{fig:cfig1bg} (a) shows the long-time intensity map and corresponding $g^{(2)}(0)$ contour. 

The $g^{(2)}(0)$ value at any location is determined by the combined $g^{(2)}(0)$ counts from: emitter-emitter, emitter-background, and background-background coincident events as in Eqn~\ref{eq:g2BG}.  Hence, in the zero background cases for three emitters, we observe the expected value of 0.67 for $g^{(2)}(0)$ when measuring equidistant from three equal brightness emitters. Background increases the value of $g^{(2)}(0)$ due to the added photon counts, and hence coincidences, from the background emitters. Additionally, the inclusion of background leads to a signal increase towards 1 further away from the emitters, as background counts become the primary source of $g^{(2)}(0)$ counts. 

Figure~\ref{fig:cfig1bg} (b) shows $w_{\text{eff}}$ as a function of measurement time. The $w_{\text{eff}}$ scaling begins with $m_1 = -0.11 \pm 0.07$ until $t_{\text{knee}} = 9 \times 10^{3}$ $\pm$ $2 \times 10^{3}P_{i,0}t$ where we transition to $m_2 = -0.505 \pm 0.006$. This transition occurs when $w_{\text{eff}}$ approximately becomes less than $d_{\text{min}} = 0.3354\sigma$.

Figure~\ref{fig:cfig1bg} (c) shows $w_{\text{eff}}$ with varied weighting parameter, $\alpha$. With optimal weighting to minimize $w_{\text{eff}}$, we obtain a superresolution factor of $\gamma$ =  8 $\pm$ 1 at $\alpha = 0.1$, $\gamma$ = 360 $\pm$ 10 at $\alpha = 0.1$, and $\gamma$ = $1.60\times 10^{5}$ $\pm$ 2 $\times 10^{3}$ at $\alpha = 0.1$, for times $100P_{i,0}t$, $5.456 \times 10^{6}P_{i,0}t$, and $10^{12}P_{i,0}t$, respectively. 

Figure~\ref{fig:cfig1bg} (d) shows the AIC goodness of fit given the data as a function of time. For this configuration, $L_{\text{AIC}} = 1$ for the two emitter model at time $100P_{i,0}t$, with a 0.63 goodness of fit score for the three emitter model. For all subsequent times, $L_{\text{AIC}} = 1$ for the three emitter model, with relatively low goodness of fit scores for all other models as we approach the long time limit.

\subsection{Four emitters of equal brightness with constant background}

For this configuration, we have added an extra emitter on a field of constant background and have also increased the average spacing of the emitters. As we have noticed that the inclusion of background increases the $g^{(2)}(0)$ values obtained further away from the emitters, we will study if this leads to changes in AIC and $w_{\text{eff}}$ scaling results.

\begin{figure}[tb!]
    \centering
    \includegraphics[width = 0.9\columnwidth]{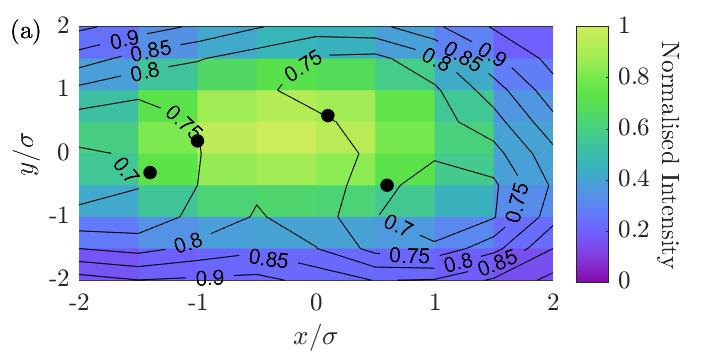}
    \includegraphics[width = 0.9\columnwidth]{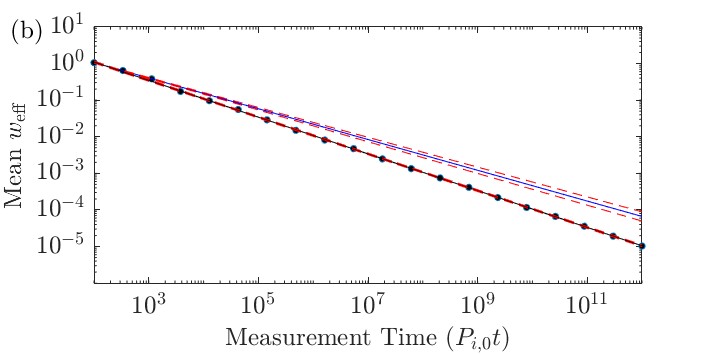}
    \includegraphics[width = 0.9\columnwidth]{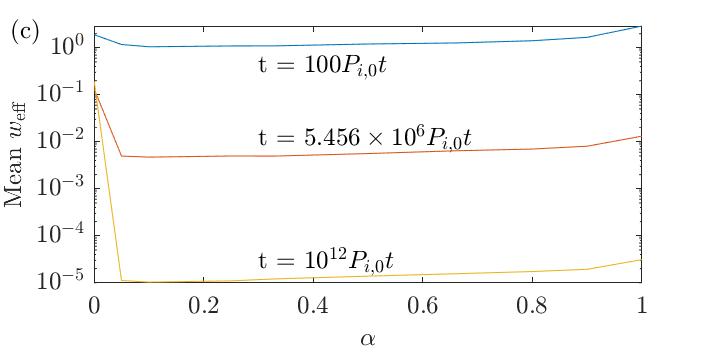}
    \includegraphics[width = 0.9\columnwidth]{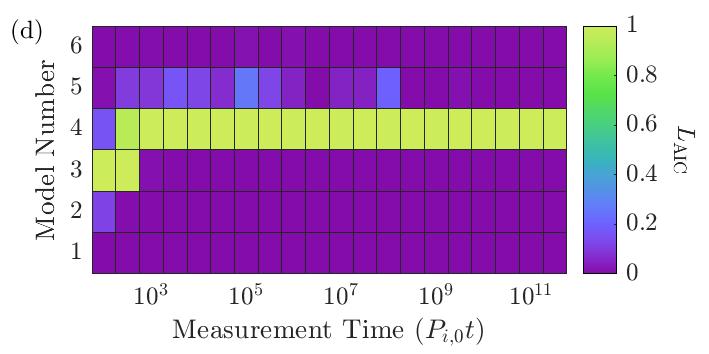}
    \caption{(a) Intensity plot and $g^{(2)}(0)$ contour of 4 emitters on a field with background $0.2P_{i,0}$ at positions: $(x_1,y_1) =(-1,0.2)\sigma$, $(x_2,y_2) = (-1.4,-0.3)\sigma$, $(x_3,y_3) = (0.6,-0.5)\sigma$, and $(x_4,y_4) = (0.1,0.6)\sigma$. Black dots indicate ground truth emitter position. (b) $w_{\text{eff}}$ scaling with measurement time. Blue line shows fitted data belonging to $w_{\text{eff},1}$ used to obtain $m_1$. Black line shows fitted data belonging to $w_{\text{eff},2}$ used to obtain $m_2$. Red dashed lines show $95\%$ confidence interval of fits. $t_\text{knee}$ is 200 $\pm$ 300 $P_{i,0}t$. (c) Average of all emitter $w_{\text{eff}}$ achieved with different intensity to $g^{(2)}(0)$ weighting ratios, $\alpha$. There is an improved localisation at higher $\alpha$ compared to the four emitter, no background case (Fig.~\ref{fig:cfig2} (c)). (d) AIC likelihood of fit for models with differing numbers of emitters. Best model has a score of 1, with others having score that shows the likelihood of model being an alternative fit to data. Similar to Figure~\ref{fig:cfig1bg} (d), we see more consistency in the optimal model at lower time. The optimal model is also determined at a lower time than in the zero background case (Fig.~\ref{fig:cfig2} (d)).}
    \label{fig:cfig4bg}
\end{figure}

Figure~\ref{fig:cfig4bg} (a) shows the intensity map of four, equal brightness emitters in a field with a background of $0.2P_{i,0}$ that would be obtained after a measurement time of $10^{12}P_{i,0}t$. Overlapping the intensity map is the corresponding $g^{(2)}(0)$. The emitters are located at positions: $(x_1,y1) =(-1,0.2)$, $(x_2,y_2) = (-1.4,-0.3)$, $(x_3,y_3) = (0.6,-0.5)$, and $(x_4,y_4) = (0.1,0.6)$. Unlike in the previous cases, the average spacing of the emitters is greater than $\sigma$, with a spacing of $1.420\sigma$. However, the minimum spacing of the emitters in this case is still less than $\sigma$, else this configuration would already be fully resolvable. 

As in the previous case, we observe areas where $g^{(2)}(0)$ exceeds the expected maximum, 0.75, for a configuration of four equal brightness emitters. This can be attributed to the inclusion of background related coincidence events. We observe closed loop regions of $g^{(2)}(0)$ where the signal reduces further away from the emitters, before increasing towards 1 as we move to the edges of the field of view. This region is a transition area where the primary source of $g^{(2)}(0)$ signal transitions from being mostly emitter coincidences to background related coincidences.

Figure~\ref{fig:cfig4bg} (b) shows the $w_{\text{eff}}$ scaling with respect to measurement time. The resolution scaling begins with $m_1 = -0.42 \pm 0.08$ until approximately $t_{\text{knee}} = 200 \pm 300 P_{i,0}t$ where we transition to $m_2 = -0.502 \pm 0.002$. This transition occurs at approximately where $w_{\text{eff}}$ becomes less than $d_{\text{min}} = 0.6403\sigma$.

Figure~\ref{fig:cfig4bg} (c) shows the calculated $w_{\text{eff}}$ with varied weighting parameter, $\alpha$. With optimal weighting to minimize $w_{\text{eff}}$, we obtain a superresolution factor of $\gamma$ = 4 $\pm$ 2 at $\alpha = 0.1$, $\gamma$ = 845 $\pm$ 8 at $\alpha = 0.1$, and $\gamma$ = 3.87 $\times 10^{5}$ $\pm$ 2 $\times 10^{3}$ at $\alpha = 0.1$, for times $100P_{i,0}t$, $5.456 \times 10^{6}P_{i,0}t$ and $10^{12}P_{i,0}t$, respectively.

Figure~\ref{fig:cfig4bg} (d) shows the AIC goodness of fit given the data as a function of time.  For this configuration, $L_{\text{AIC}} = 1$ for the three emitter model at times $100P_{i,0}t$, and $336P_{i,0}t$. However, at time $336P_{i,0}t$ the four emitter model has a goodness of fit score of 0.94, making it almost equally viable as a fit to the three emitter model. For all subsequent times, $L_{\text{AIC}} = 1$ for the three emitter model, with relatively low goodness of fit scores for all other models as we approach the long time limit.

\subsection{Discussion on constant background}\label{sect:ConstBack}

The scaling of our configurations in Figures~\ref{fig:cfig1bg} \& \ref{fig:cfig4bg} behave in the same manner as cases studied with zero background and keeps with the heuristic developed in Section~\ref{sect:FourLine}. $t_\text{knee}$ is observed approximately where $w_{\text{eff}}$ becomes less than $d_\text{min}$.

We observe that $w_{\text{eff}}$ is minimised at $\alpha$ values higher than in the zero background cases (Figs~.\ref{fig:cfig4} \& Figs~.\ref{fig:cfig2}) for times $100P_{i,0}t$, $5.456 \times 10^{6}P_{i,0}t$, and $10^{12}P_{i,0}t$, with optimal $\alpha$ being at 0.1 compared to 0.05. This together with the observation that, for cases with background, AIC converges to the ground truth model faster than in the zero background cases, suggests that background is assisting in localisation. From the $w_{\text{eff}}$ results in Figures ~\ref{fig:cfig1bg} (c) \& \ref{fig:cfig4bg} (c), we see that high values of $\alpha$, as well as $\alpha = 1$ perform noticeably better than the zero background cases, particularly at time $10^{12}P_{i,0}t$ where $\alpha = 1$ was at the order of, or worse than intensity localisation alone. From this, we can infer that the improvement in localisation that comes with background is caused by increased coincidence counts occurring from background emitters.

\subsection{Effects of increasing background}

To determine the effect of increasing background on $w_{\text{eff}}$, we consider two four-emitter configurations cases where we vary the background levels. These cases are summarised in Table~\ref{BGtable}.

\begin{table*}[tb]
\caption{Summary for cases in Figs.~\ref{fig:BGscale2} and \ref{fig:BGscale4}. $m_2$ and $t_\text{knee}$ are calculated in each case for background levels ranging from $0.002P_{i,0}$ to $10P_{i,0}$. For cases where $t_\text{knee}$ can not be interpolated, from $m_1$ and $m_2$, $t_\text{knee}$ is given as the first data point where $w_\text{eff} < d_\text{min}$ and is marked by *.}
\begin{center}
\begin{tabular}{| c | p{2.6cm} | p{2.6cm} | p{2.6cm} | p{2.6cm} |  p{2.6cm} |}
\hline
\hline
$(x_i,y_i)/\sigma$ & \multicolumn{2}{c|}{close} & \multicolumn{2}{c|}{far} \\ \cline{2-5}
 & \multicolumn{2}{c|}{$(-0.15,-0.3)$, $(-0.45,0)$, $(0.1,1)$, $(0.15,0.1$)} & \multicolumn{2}{c|}{$(-1,0.2)$, $(-1.4,-3)$, $(0.6,-0.5)$, $(0.1,0.6)$} \\ \hline

 Background Level~[$P_{bg}$] & $m_2$  & $t_\text{knee}~[P_{i,0}t$]  & $m_2$  & $t_\text{knee}~[P_{i,0}t$] \\
 \hline \hline
0.002$P_{i,0}$ & -0.50 $\pm$ 0.01 & 4.2 $\times$ $10^{4}$ $\pm$ 2 $\times$ $10^{3}$ & -0.500 $\pm$ 0.009 & 336 * \\ \hline
0.02$P_{i,0}$ & -0.49 $\pm$ 0.01 & 2.2 $\times$ $10^{4}$ $\pm$ 2 $\times$ $10^{3}$ & -0.500 $\pm$ 0.004 & 1.23 $\times$ $10^{3}$ * \\ \hline
0.2$P_{i,0}$ & -0.50 $\pm$ 0.01 & 7 $\times$ $10^{3}$ $\pm$ 4 $\times$ $10^{3}$ & -0.504 $\pm$ 0.003 & 1.23 $\times$ $10^{3}$ * \\ \hline
0.5$P_{i,0}$ & -0.49 $\pm$ 0.02 & 7 $\times$ $10^{3}$ $\pm$ 3 $\times$ $10^{3}$ & -0.507 $\pm$ 0.007 & 336 * \\ \hline
$P_{i,0}$ & -0.48 $\pm$ 0.02 & 4 $\times$ $10^{3}$ $\pm$ 2 $\times$ $10^{3}$ & -0.499 $\pm$ 0.002 & 1.23 $\times$ $10^{3}$ * \\ \hline
2$P_{i,0}$ & -0.47 $\pm$ 0.01 & 4 $\times$ $10^{3}$ $\pm$ 1 $\times$ $10^{3}$ & -0.501 $\pm$ 0.003 & 1.23 $\times$ $10^{3}$ * \\ \hline
10$P_{i,0}$ & 0.00 $\pm$ 0.02 & 1.28 $\times$ $10^{4}$ * & -0.46 $\pm$ 0.03 & 3 $\times$ $10^{3}$ $\pm$ 8 $\times$ $10^{3}$ \\ \hline
\hline
\end{tabular}
\end{center}
\label{BGtable}
\end{table*}

\begin{figure}[tb!]
    \centering
    \includegraphics[width = 0.9\columnwidth]{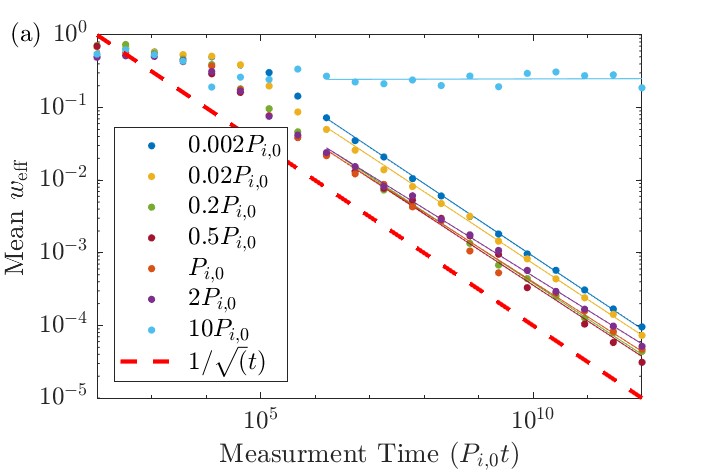}
    \includegraphics[width = 0.9\columnwidth]{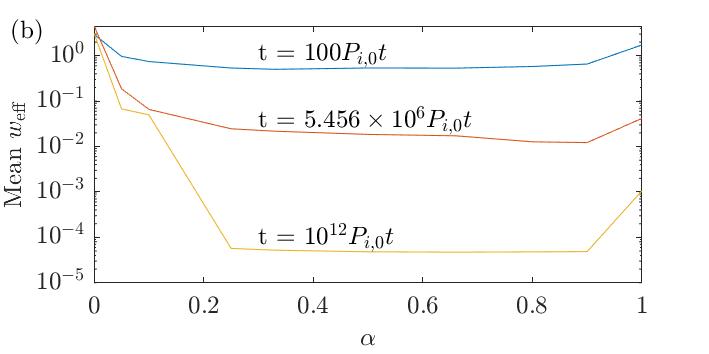}
    \caption{(a) $w_{\text{eff}}$ scaling with time for a configuration with four emitters located at $(x_1,y1) =(-0.15,-0.3)$, $(x_2,y_2) = (-0.45,0)$, $(x_3,y_3) = (0.1,1)$, and $(x_4,y_4) = (0.15,0.1)$, with a background varied from $P_{bg} = 0.002P_{i,0}$ to $P_{bg} = 10P_{i,0}$. Red-dashed line indicates $1/\sqrt{t}$ $w_{\text{eff}}$ scaling, which is the best expected scaling result. (b) $w_{\text{eff}}$ for varying $g^{(2)}(0)$ weightings, $,\alpha$, at a background of $P_{bg} = P_{i,0}$. Note that $w_{\text{eff}}$ is minimized at higher $\alpha$ compared to cases studied at zero background.}
    \label{fig:BGscale2}
\end{figure}

Figure~\ref{fig:BGscale2} (a) shows localisation as a function of measurement time for a configuration of four equal brightness emitters, with average spacing $0.59\sigma$, i.e. sub-diffraction limit, for increasing  background. The emitters are located at positions: $(x_1,y1) =(-0.15,-0.3)\sigma$, $(x_2,y_2) = (-0.45,0)\sigma$, $(x_3,y_3) = (0.1,1)\sigma$, and $(x_4,y_4) = (0.15,0.1\sigma)$.  

We observe from Table~\ref{BGtable} that, for the configuration in Figure~\ref{fig:BGscale2}, $t_\text{knee}$ reduces as we increase background from $0.002P_{i,0}$ to $2P_{i,0}$. As we have observed in the cases from Figures~\ref{fig:cfig1bg} \& \ref{fig:cfig4bg}, background coincident counts are aiding in localisation. Looking at Figure~\ref{fig:BGscale2} (b), which shows the $\alpha$ weighing at a background level of $P_{i,0}$, we see that a shift favouring higher values of $\alpha$ has occurred, even compared to the the results seen in Figures~\ref{fig:cfig1bg} \& \ref{fig:cfig4bg}, which were at a background of $0.2P_{i,0}$. Here, $w_{\text{eff}}$ is minimised at $\alpha = 0.33$, $\alpha = 0.90$, and $\alpha = 0.66$ for times $100P_{i,0}t$, $5.456 \times 10^{6}P_{i,0}t$ and $10^{12}P_{i,0}t$, respectively. As suggested in Section~\ref{sect:ConstBack}, this is further evidence that background-emitter and background-background correlations can aid in localisation.

While increasing the amount of $g^{(2)}(0)$ counts by increasing background levels does aid in localisation, we can also see from Figure~\ref{fig:BGscale2} and Table~\ref{BGtable} that there appears to be a limit, as seen for background level $10P_{i,0}$. While we do eventually reach a point where $w_{\text{eff}} < d_{\text{min}}$, which occurs some time around the data point at 1.28 $\times$ $10^{4}P_{i,0}t$, there is negligible change in $w_{\text{eff}}$ from this point as the algorithm is unable to localise any more precisely. At this extreme background, we expect that at higher times, the background-background correlations will mostly overpower any emitter related correlations preventing further localisation.

\begin{figure}[tb!]
    \centering
    \includegraphics[width = 0.9\columnwidth]{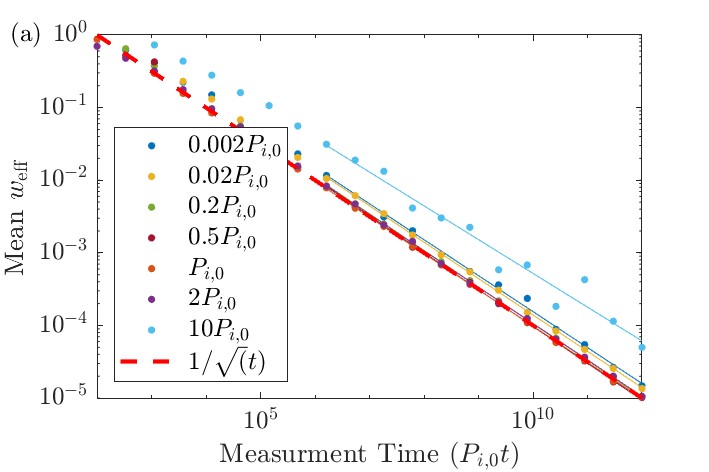}
    \includegraphics[width = 0.9\columnwidth]{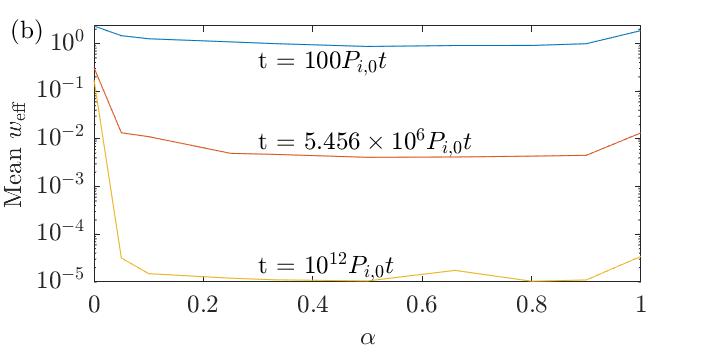}
    \caption{(a) $w_{\text{eff}}$ scaling with time for a configuration with four emitters located at $(x_1,y1) =(-1,0.2)$, $(x_2,y_2) = (-1.4,-3)$, $(x_3,y_3) = (0.6,-0.5)$, and $(x_4,y_4) = (0.1,0.6)$, with a background ranging from $P_{bg} = 0.002P_{i,0}$ to $P_{bg} = 10P_{i,0}$. Due to larger emitter spacing, the effects of high background is not as significant as in Figure~\ref{fig:BGscale2}. Red-dashed line indicates $1/\sqrt{t}$ $w_{\text{eff}}$ scaling, which is the best expected scaling result. (b) $w_{\text{eff}}$ results at various $g^{(2)}(0)$ weightings, $\alpha$, with a background of $P_{bg} = P_{i,0}$. $w_{\text{eff}}$ is slightly more minimized at higher $\alpha$ weighting compared to cases studied at zero background.}
    \label{fig:BGscale4}
\end{figure}

Another case where background is increased in the same increments as in Figure~\ref{fig:BGscale2} is shown in Figure~\ref{fig:BGscale4} (a) for a configuration of four equal brightness emitters with an average spacing greater than $\sigma$ (1.42$\sigma$). The emitters are located at positions: $(x_1,y1) =(-1,0.2)\sigma$, $(x_2,y_2) = (-1.4,-3)\sigma$, $(x_3,y_3) = (0.6,-0.5)\sigma$, and $(x_4,y_4) = (0.1,0.6)\sigma$.

Unlike in the previous case, there is little change in $t_\text{knee}$ between background levels until we reach $10P_{i,0}$. This can be seen from Table~\ref{BGtable} and Figure~\ref{fig:BGscale4} (a), where the $w_{\text{eff}}$ lines occupy mostly the same region. Background level $10P_{i,0}$ is a noticeable exception, and though we achieve an $m_2$ of close to $1/\sqrt{t}$, there is a higher uncertainty relative to lower background levels, and $t_\text{knee}$ has a considerably high uncertainty due to fluctuations in $w_{\text{eff}}$ in the long time region. Our ability to resolve the particles despite the high background can be attributed to the spacing of the emitters around the field of view, which allows for more regions where $g^{(2)}(0)$ does not immediately go to 1 due to background-background coincidences.

 As in Figure~\ref{fig:BGscale2} (b), in Figure~\ref{fig:BGscale4} (b) we see $w_{\text{eff}}$ minimisation take place at higher $\alpha$ when at background level $P_{bg} = P_{i,0}$ compared to zero background and lower background cases such as Figures~\ref{fig:cfig2} \& \ref{fig:cfig4bg}. Here, $w_{\text{eff}}$ is minimised at $\alpha = 0.60$, $\alpha = 0.60$, and $\alpha = 0.80$ for times $100P_{i,0}t$, $5.456 \times 10^{6}P_{i,0}t$ and $10^{12}P_{i,0}t$, respectively.

\begin{figure}[tb!]
    \centering
    \includegraphics[width = 0.9\columnwidth]{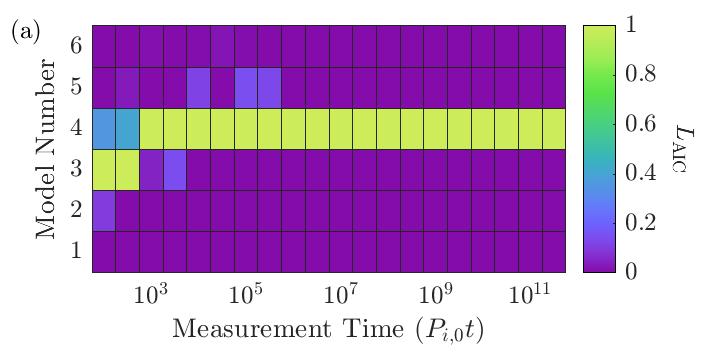}
    \includegraphics[width = 0.9\columnwidth]{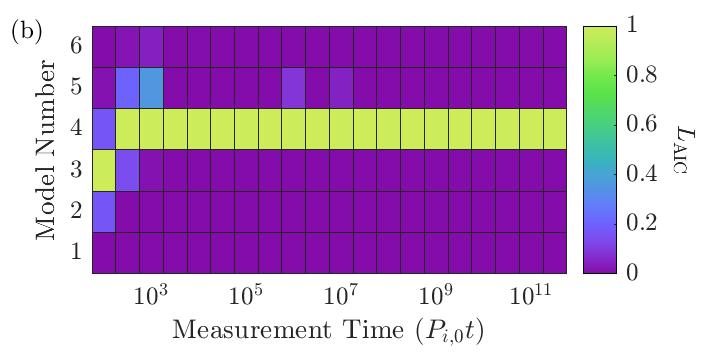}
    \includegraphics[width = 0.9\columnwidth]{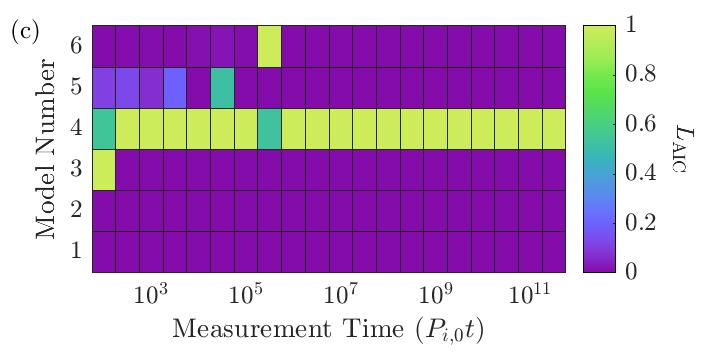}
    \caption{AIC results at background levels: $P_{bg} = 0.002P_{i,0}$, $P_{bg} = 0.2P_{i,0}$, and $P_{bg} = P_{i,0}$ for configuration studied in Figure~\ref{fig:BGscale2}. The time required for AIC to converge to the ground truth model (Model Number 4 in this case), reduces slightly as background increases.}
    \label{fig:BGAICclose}
\end{figure}

\begin{figure}[tb!]
    \centering
    \includegraphics[width = 0.9\columnwidth]{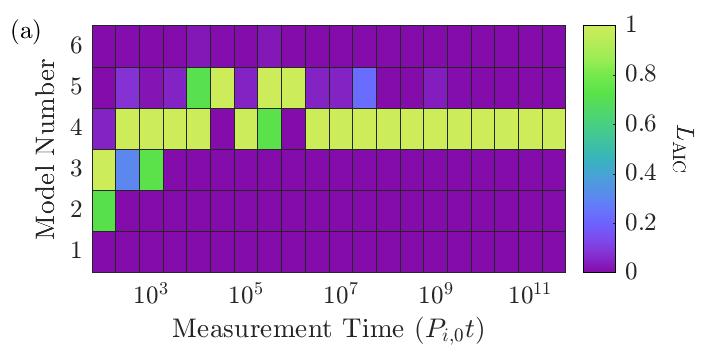}
    \includegraphics[width = 0.9\columnwidth]{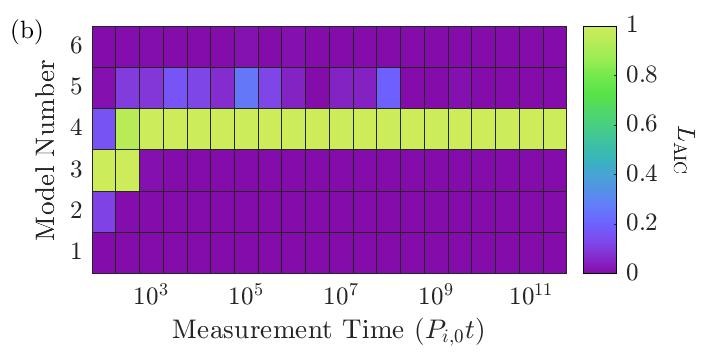}
    \includegraphics[width = 0.9\columnwidth]{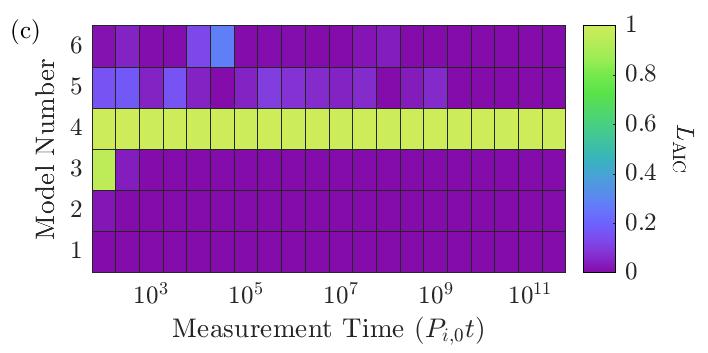}
    \caption{AIC results at background levels: $P_{bg} = 0.002P_{i,0}$, $P_{bg} = 0.2P_{i,0}$, and $P_{bg} = P_{i,0}$ for configuration studied in Figure~\ref{fig:BGscale4}. As in Figure~\ref{fig:BGscale4}, an increase in background leads to a minor reduction in the time required for the optimal model to converge to Model Number 4, which is the number of ground truth emitters.}
    \label{fig:BGAICfar}
\end{figure}

We study the effects of background on AIC for the cases shown in Figures~ \ref{fig:BGscale2} and \ref{fig:BGscale4} by determining AIC at background levels: $P_{bg} = 0.002P_{i,0}$, $P_{bg} = 0.2P_{i,0}$, and $P_{bg} = P_{i,0}$. (Figures~\ref{fig:BGAICclose} and \ref{fig:BGAICfar}). 

For the close emitter case in Figure~\ref{fig:BGAICclose}, we see that the model resulting in $L_{\text{AIC}} = 1$ is consistent through to the long time limit, with a slight decrease in the time required for model number 4 for be the optimal model as we increase background levels. For the far emitter case in Figure~\ref{fig:BGAICfar}, we see a increase in consistency for the optimal model at higher background levels, and once again observe a slight decrease in time required for model number 4 to be the optimal model. This suggests that background coincidence counts can contribute information that the algorithm can use to constrain emitter number, which can also play a part in improving localisation precision.

\section{Study of Unequal Brightness Emitter Configurations}

So far, we have studied the simple case of equal brightness emitters, i.e., $P_{i,0} = P_{j,0}$. To show that this approach can be applied to the general case, and to determine if this results in a change in scaling behavior, we will now study fields of emitters of unequal brightness'. We will consider the cases of no background, and with background.

\subsection{Four emitters of unequal brightness with no background}

\begin{figure}[tb!]
    \centering
    \includegraphics[width = 0.9\columnwidth]{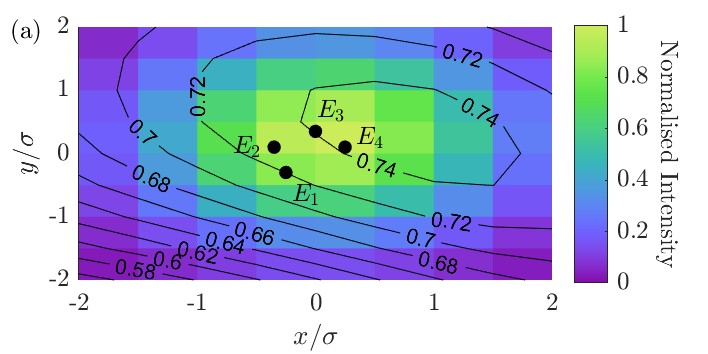}
    \includegraphics[width = 0.9\columnwidth]{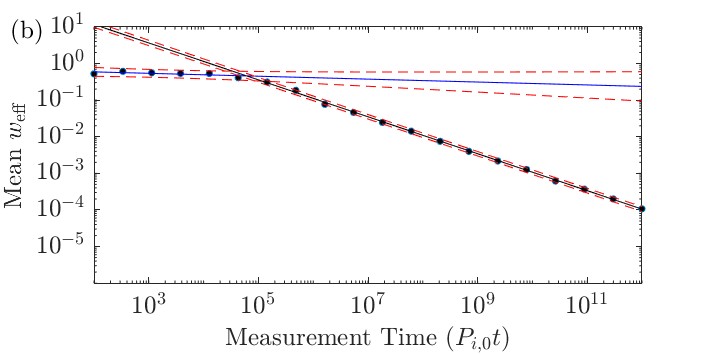}
    \includegraphics[width = 0.9\columnwidth]{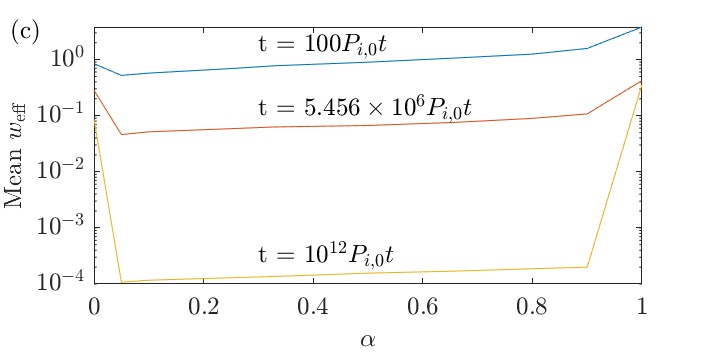}
    \includegraphics[width = 0.9\columnwidth]{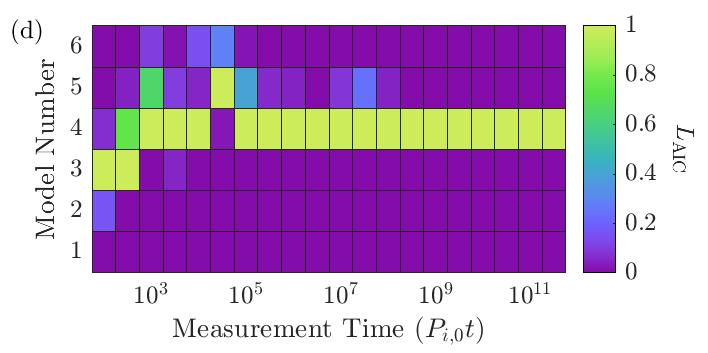}
    \caption{(a) Intensity plot and $g^{(2)}(0)$ contour of 4 emitters on a field with no background at positions: $(x_1,y_1) =(-0.25,-0.3)\sigma$, $(x_2,y_2) = (-0.35,0.1)\sigma$, $(x_3,y_3) = (0,0.35)\sigma$, and $(x_4,y_4) = (0.25,0.1)\sigma$, with relative brightness': $P_{2,0} = 0.83P_{1,0}$, $P_{3,0} = 0.67P_{1,0}$, and $P_{4,0} = 0.5P_{1,0}$. Black dots indicate ground truth emitter position. (b) $w_{\text{eff}}$ scaling with measurement time. Blue line shows fitted data belonging to $w_{\text{eff},1}$ used to obtain $m_1$. Black line shows fitted data belonging to $w_{\text{eff},2}$ used to obtain $m_2$. Red dashed lines show $95\%$ confidence interval of fits. $t_\text{knee}$ is $7 \times 10^{4} \pm 1 \times 10^{4}P_{i,0}t$. (c) Average of all emitter $w_{\text{eff}}$ achieved with different intensity to $g^{(2)}(0)$ weighting ratios, $\alpha$. (d) AIC goodness of fit given the data for models with differing numbers of emitters. Best model has a score of 1, with others having a score showing the goodness of fit relative to the optimal model.}
    \label{fig:cfig5UQ}
\end{figure}

Figure~\ref{fig:cfig5UQ} (a) shows the intensity map of four, unequal brightness emitters on a field with no background that would be obtained after an infinitely long measurement time. The emitters have relative brightness $P_{2,0} = 0.83P_{1,0}$, $P_{3,0} = 0.67P_{1,0}$, and $P_{4,0} = 0.5P_{1,0}$. Overlapping the intensity map is the corresponding $g^{(2)}(0)$ contour. The emitters are located at positions: $(x_1,y1) =(-0.25,-0.3)\sigma$, $(x_2,y_2) = (-0.35,0.1)\sigma$, $(x_3,y_3) = (0,0.35)\sigma$, and $(x_4,y_4) = (0.25,0.1)\sigma$.

By comparing Figure~\ref{fig:cfig5UQ} (a) to the previous equal-brightness no-background cases, we can see that while the general $g^{(2)}(0)$ contour geometry resembles the equal-brightness no-background case, a shift has occurred in the location of the highest $g^{(2)}(0)$ value. The highest $g^{(2)}(0)$ value region corresponds to the area where the received brightness' of all emitters are equal, which for the equal brightness case, is in the middle and equidistant from all emitters. Here, a shift has occurred in the direction of the weaker emitters because the change in brightness further away from the emitters is no longer the same for every emitter.

Figure~\ref{fig:cfig5UQ} (b) shows the $w_{\text{eff}}$ scaling with respect to measurement time. The $w_{\text{eff}}$ scaling begins at $m_1 = -0.04 \pm 0.06$, until approximately $t_{\text{knee}} = 7 \times 10^{4} \pm 1 \times 10^{4}P_{i,0}t$, where we transition to  $m_2 = -0.504 \pm 0.008$. This transition occurs where $w_{\text{eff}}$ approximately becomes less than $d_{\text{min}} = 0.3536\sigma$, following the same trend as in the equal brightness cases. The relationship between $t_{\text{knee}}$, $w_{\text{eff}}$ and $d_{\text{min}}$ has remained unchanged from the equal brightness cases.

In Figure~\ref{fig:cfig5UQ} (c), we show the calculated $w_{\text{eff}}$ with varying $\alpha$. With optimal weighting to minimize $w_{\text{eff}}$, we obtain a superresolution factor of $\gamma$ = 8 $\pm$ 4 at $\alpha = 0.05$, $\gamma$ = 88 $\pm$ 4 at $\alpha = 0.05$, and $\gamma$ = 3.7 $\times 10^{4}$ $\pm$ 600 at $\alpha = 0.05$, for times $100P_{i,0}t$, $5.456 \times 10^{6}P_{i,0}t$, and $10^{12}P_{i,0}t$, respectively.

Figure~\ref{fig:cfig5UQ} (d) shows the AIC goodness of fit given the data as a function of time. For this configuration, the model resulting in $L_{\text{AIC}} = 1$ varies between three and five until approximately $P_{1,0}t = 1.438 \times 10^{5}P_{i,0}t$, where all subsequent optimal models are the are four emitter models.

Compared to the previous equal-brightness no-background cases (Figs.~\ref{fig:cfig4} \& \ref{fig:cfig2}), both weighting and AIC behaviour appear unchanged, with $\alpha$ plateauing between 0 and 1, and AIC results converging on the model matching the ground truth as time increases.

\subsection{Four emitters of unequal brightness with constant background}

\begin{figure}[tb!]
    \centering
    \includegraphics[width = 0.9\columnwidth]{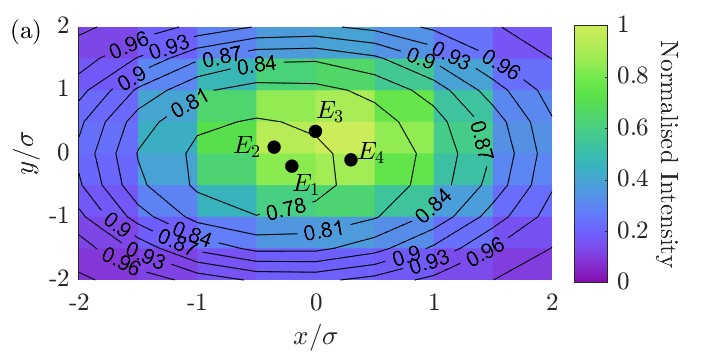}
    \includegraphics[width = 0.9\columnwidth]{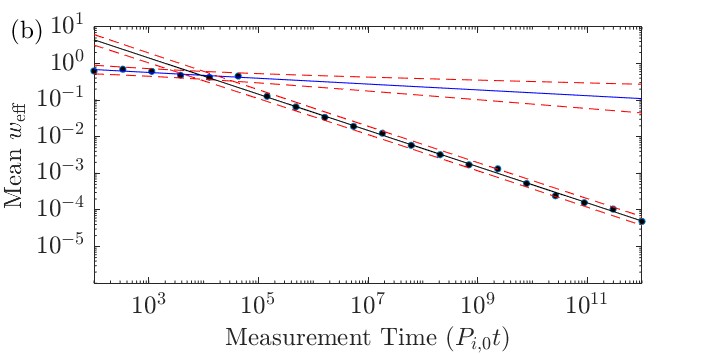}
    \includegraphics[width = 0.9\columnwidth]{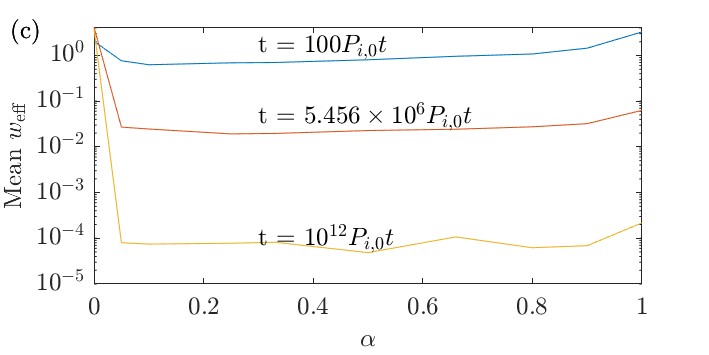}
    \includegraphics[width = 0.9\columnwidth]{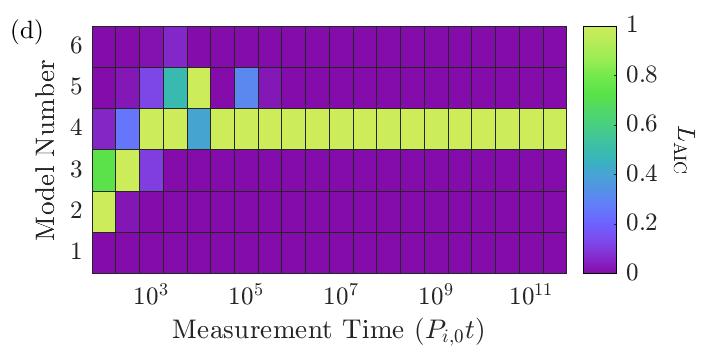}
    \caption{(a) Intensity plot and $g^{(2)}(0)$ contour of 4 emitters on a field with background level: $0.2P_{1,0}$, at positions: $(x_1,y1) =(-0.2,-0.2)\sigma$, $(x_2,y_2) = (-0.35,0.1)\sigma$, $(x_3,y_3) = (0,0.35)\sigma$, and $(x_4,y_4) = (0.3,-0.1)\sigma$, with relative brightness': $P_{2,0} = 0.83P_{1,0}$, $P_{3,0} = 0.67P_{1,0}$, and $P_{4,0} = 0.5P_{1,0}$. Black dots indicate ground truth emitter position. (b) $w_{\text{eff}}$ scaling with measurement time. Blue line shows fitted data belonging to $w_{\text{eff},1}$ used to obtain $m_1$. Black line shows fitted data belonging to $w_{\text{eff},2}$ used to obtain $m_2$. Red dashed lines show $95\%$ confidence interval of fits. $t_\text{knee}$ is $1.0 \times 10^{4} \pm 3 \times 10^{3}P_{i,0}t$. (c) Average of all emitter $w_{\text{eff}}$ achieved with different intensity to $g^{(2)}(0)$ weighting ratio, $\alpha$. As in the equal brightness cases, the inclusion of background has improved the localisation performance at higher values of $\alpha$ compared to cases with no background. (d) AIC goodness of fit given the data for models with differing numbers of emitters. Best model has a score of 1, with others having a score showing the goodness of fit relative to the optimal model.}
    \label{fig:cfigbg5UQ}
\end{figure}

Figure~\ref{fig:cfigbg5UQ} (a) shows the intensity map of four, unequal brightness emitters on a field with a uniform background of brightness $0.2P_{1,0}$ that would be obtained after an infinitely long measurement time. The emitters have relative brightness' $P_{2,0} = 0.83P_{1,0}$, $P_{3,0} = 0.67P_{1,0}$, and $P_{4,0} = 0.5P_{1,0}$. Overlapping the intensity map is the corresponding $g^{(2)}(0)$ contour. The emitters are located at positions: $(x_1,y1) =(-0.2,-0.2)\sigma$, $(x_2,y_2) = (-0.35,0.1)\sigma$, $(x_3,y_3) = (0,0.35)\sigma$, and $(x_4,y_4) = (0.3,-0.1)\sigma$.

In this configuration we once again see a shift in the $g^{(2)}(0)$ contour compared to the previous equal-brightness with-background cases. However, note that the shift has occurred in the direction of the brightest emitters rather than the weakest as was seen in Figure~\ref{fig:cfig5UQ}. This shift has occurred as a consequence of introducing background, as the contributions of background-background correlations, which increase $g^{(2)}(0)$ towards 1, are more significant in areas further away from the brightest emitters. 

Figure~\ref{fig:cfigbg5UQ} (b) shows the $w_{\text{eff}}$ scaling with respect to measurement time. The $w_{\text{eff}}$ scaling begins at $m_1 = -0.08 \pm 0.06$, until approximately $t_{\text{knee}} = 1.0 \times 10^{4} \pm 3 \times 10^{3}P_{i,0}t$ where we transition to $m_2 = -0.49 \pm 0.01$. This transition occurs at approximately when $w_{\text{eff}}$ becomes less than $d_{\text{min}} = 0.3354\sigma$.

In Figure~\ref{fig:cfigbg5UQ} (c), we show the calculated $w_{\text{eff}}$ with varying $\alpha$. With optimal weighting to minimize $w_{\text{eff}}$, we obtain a superresolution factor of $\gamma$ = 6 $\pm$ 5 at $\alpha = 0.1$, $\gamma$ = 210 $\pm$ 10 at $\alpha = 0.25$, and $\gamma$ = 8.2 $\times 10^{4}$ $\pm$ 2 $\times 10^{3}$ at $\alpha = 0.5$, for times $100P_{i,0}t$, $5.456 \times 10^{6}P_{i,0}t$, and $10^{12}P_{i,0}t$, respectively. As in the previous cases with background, we see that the inclusion of background can potentially increase the value of $\alpha$ that minimises $w_{\text{eff}}$.

Figure~\ref{fig:cfig5UQ} (d) shows the AIC goodness of fit given the data as a function of time. For this configuration, the model resulting in $L_{\text{AIC}} = 1$ increases from two to five emitter models until time approximatly $1.274 \times 10^{4}P_{i,0}t$, where all subsequent optimal models are the are four emitter models. For this case, converging on the ground truth model does not occur as rapidly as in the previous equal-brightness, constant background cases with background levels at $0.2P_{1,0}$ (Figs.~ \ref{fig:cfig1bg}, \ref{fig:cfig4bg}, \ref{fig:BGAICclose}(b), \& \ref{fig:BGAICfar}(b)).

\subsection{Relationship between scaling and emitter spacing for unequal brightness cases}

The scaling behaviour of the unequal brightness cases behave in the same manner as the equal brightness cases with, and without background. We once again see that the time required to achieve $1/\sqrt{t}$ scaling, $t_\text{knee}$, is tied to $d_{\text{min}}$, and we observe two separate scaling behaviours before and after the knee. To corroborate this for various emitter spacing's, and compare correlation-with-intensity to intensity-only localisation for unequal brightness cases, we consider the scaling of additional cases as done in Section~\ref{sect:EQscale}.

\begin{figure}[tb!]
    \centering
    \includegraphics[width = 0.9\columnwidth]{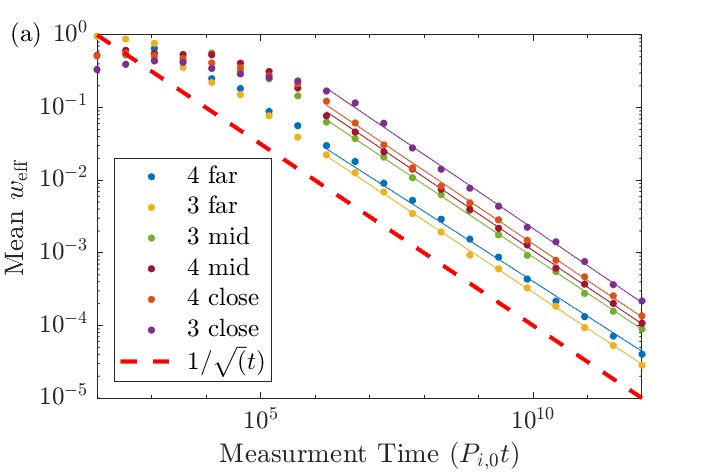}
    \includegraphics[width = 0.9\columnwidth]{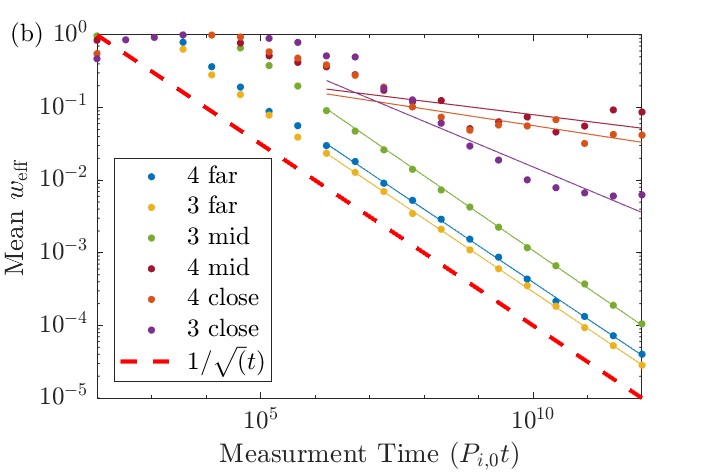}
    \caption{(a) Localisation scaling with time of six different emitter configurations categorised by their minimum emitter spacing relative to other configurations with the same emitter number, with "close" having the smallest spacing, and "far" being the furthest. (b) Resolution scaling for the same configurations in (a), using only intensity information. From top to bottom according to the legend, the minimum spacing for each configuration is: $0.8062\sigma$, $0.6103\sigma$, $0.4031\sigma$, $0.3536\sigma$, $0.2693\sigma$, and $0.1803\sigma$. The emitters have relative brightness': $P_{2,0} = 0.83P_{1,0}$, $P_{3,0} = 0.67P_{1,0}$, and $P_{4,0} = 0.5P_{1,0}$ for all four emitter cases, and relative brightness': $P_{2,0} = 0.75P_{1,0}$, and $P_{3,0} = 0.5P_{1,0}$ for all three emitter cases. Red-dashed line indicates $1/\sqrt{t}$ $w_{\text{eff}}$ scaling, which is the best expected scaling result.} 
    \label{fig:UQscale}
\end{figure}

In Figure~\ref{fig:UQscale}, we show a comparison of the $w_{\text{eff}}$ scaling results with respect to time for several unequal brightness emitter cases, sorted by the cases relative minimum emitter closeness and emitter number. These cases are summarised in Table~\ref{UQtable}. 

As seen in the equal brightness cases (Figure~\ref{fig:EQscale}), we observe a increase in $t_{\text{knee}}$ as $d_\text{min}$ decreases. $t_{\text{knee}}$ is higher when using intensity-only compared to intensity-with-correlation localisation, and tends to have high relative uncertainty when $t_\text{knee}$ is able to be interpolated. Additionally, we see that $m_2$ is unable to achieve $1/\sqrt{t}$ when using intensity-only for our 4 mid, 4 close, and 3 close cases.

\begin{table*}[tb!]
\caption{Summary of configurations shown in Figure~\ref{fig:UQscale}. Configurations are categorised as 'far', 'mid', or 'close' based on their minimum emitter spacing relative to other configurations with the same number of emitters, $N$. Results obtained using $g^{(2)}(0)$-with-intensity $(I$ \& $C)$ and intensity-only $(I)$ fitting are compared. Results are in order of the emitter's minimum spacing, $d_\text{min}$. For cases where the interpolated $t_{\text{knee}}$ is $< 0$, the first data point where $w_{\text{weff}} < d_{\text{min}}$ is used and marked with *. If $w_{\text{weff}}$ is $< d_{\text{min}}$ from the first data point, or $m_1$ can not be obtained accurately due to too few data points, N/A is used for the corresponding data. As was seen in Table~\ref{EQtable}, the results here follow the same trend of $t_{\text{knee}}$ increasing as $d_\text{min}$ decreases, with higher $t_{\text{knee}}$ when using intensity only, particularly at small $d_\text{min}$.}
\begin{center}
\begin{tabular}{|c|c| p{2cm} | p{2cm} | p{2cm} | p{2cm} | p{2cm} | c | p{2cm} | p{2cm}|}
\hline
\hline
$N$ & Separation & $(x_{i},y_{i})/\sigma$ & $m_1$ $(I$ \& $C)$ & $m_2$ $(I$ \& $C)$ & $m_1$ $(I)$ & $m_2$ $(I)$ & $d_{\text{min}}~[\sigma]$ & $t_\text{knee}~[P_{i,0}t$] $(I$ \& $C)$ & $t_\text{knee}~[P_{i,0}t$] ($I$) \\
 \hline \hline
4 & far& $(-1.0,0.3)$, $(-1.3,-0.4)$, $(0.1,-0.2)$, $(0.3,0.6)$ & N/A & $-0.48 \pm 0.01$ & $-0.270 \pm 0.008$ & $-0.503 \pm 0.007$ & 0.81 & 1.128 $\times$ $10^{3}$* & 1000 $\pm$ 100\\ \hline
3 & far & $(0,-0.4)$, $(-0.5,-0.5)$, $(0.5,0.4)$ & $-0.1 \pm 0.1$ & $-0.495 \pm 0.007$ & $-0.3 \pm 0.3$ & $-0.503 \pm 0.004$ & 0.61 & 1.2 $\times$ $10^{3}$ $\pm$ 500 & 600 $\pm$ 2 $\times$ $10^{3}$\\ \hline
3 & mid & $(-0.3,-0.35)$, $(0.05,-0.10)$, $(0.25,-0.45)$ & $-0.0 \pm 0.1$ & $-0.50 \pm 0.01$  & $-0.1 \pm 0.2$ & $-0.515 \pm 0.008$ & 0.40 & 3 $\times$ $10^{4}$ $\pm$ 1 $\times$ $10^{4}$ & 2 $\times$ $10^{4}$ $\pm$ 1 $\times$ $10^{4}$\\  \hline
4 & mid & $(-0.25,-0.3)$, $(-0.35,0.1)$, $(0,0.35)$, $(0.25,0.1)$ & $-0.04 \pm 0.06$ & $-0.504 \pm 0.008$ & $-0.14 \pm 0.08$ & $-0.09 \pm 0.08$ & 0.35 & 7 $\times$ $10^{4}$ $\pm$ 1 $\times$ $10^{4}$ & 5.456 $\times$ $10^{6}$*\\ \hline
4 & close & $(-0.10,-0.15)$, $(-0.3,0.05)$, $(0.05,0.25)$, $(0.15,-0.25)$ & $-0.09 \pm 0.04$ & $-0.50 \pm 0.01$ & $-0.10 \pm 0.08$ & $-0.12 \pm 0.06$ & 0.27 & 2.5 $\times$ $10^{5}$ $\pm$ 4 $\times$ $10^{4}$ & 1.833 $\times$ $10^{7}*$\\ \hline
3 & close & $(-0.1,-0.3)$, $(0.1,-0.25)$, $(0.2,-0.4)$ & $-0.06 \pm 0.05$ & $-0.51 \pm 0.01$ & $-0.07 \pm 0.09$ & $-0.3 \pm 0.1$ & 0.18 & 1.1 $\times$ $10^{6}$ $\pm$ 2 $\times$ $10^{5}$ & 6.159 $\times$ $10^{7}*$\\
\hline
\hline
\end{tabular}
\end{center}
\label{UQtable}
\end{table*}

\section{Conclusion}

Our results show  that combining intensity and $g^{(2)}(0)$ provides improved  localisation of few single photon emitters relative to that obtained through intensity information alone. Except in the limit of very large background, where the background intensity is much greater than the emitter brightness, we observe diffraction unlimited localisation that asymptotically scales as $1/\sqrt{t}$ where $t$ is the total measurement time. Localisation scaling of $1/\sqrt{t}$ can be achieved in the limit of high background when the spacing of the emitters is close to the optical point spread function's standard deviation. Prior to achieving $1/\sqrt{t}$ scaling, all of the configurations we tested showed a weaker localisation scaling with time, which held until the protocol had isolated the emitters and then localisation was more rapid.  We term the point where the scaling laws shift as the `knee' and observe that $t_{\text{knee}}$ depends on the exact geometry of the system under consideration.

While our results provide a simple heuristic to predict scaling behaviour based on $t_{\text{knee}}$ and the geometry of the configuration, further work is required to be able to understand the scaling behaviour prior to the knee, as this could be used to predict the measurement time required to achieve $1/\sqrt{t}$, and therefore localise individual emitters, when the scale and geometry of the sample is somewhat known. It is probable that in our model, the scaling prior to the knee is a consequence of fitting multiple Gaussians to the data when the emitters are unresolved and/or the number of emitters are unknown. Our results demonstrate that the Akaike Information Criteria can be used to constrain the number of emitters. However, high measurement times are required to converge on the ground truth model in cases with very low or no background.

With the current model, we predict that this quantum correlation technique could be applied in a widefield approach when using an array of detectors, or combined with other superresolution techniques in order to improve resolution by an order of magnitude which may aid in the imaging of samples where the amount of light used must be considered, such as in bioimaging.

\section{Acknowledgments}

The authors acknowledge the assistance of Josef Worboys, Daniel Drumm, Brant Gibson, Brett Johnson, and members of the RMIT Center of Excellence for Nanoscale BioPhotonics. This work is funded by the Air Force Office of Scientific Research (FA9550-20-1-0276). ADG also acknowledges funding from the Australian Research Council (CE140100003).

\newpage

\bibliography{References}

\end{document}